\documentclass[dvips,10pt]{article}

\usepackage{a4wide,amsmath,amssymb,epsf,epsfig,amsthm,bm,multicol}

\theoremstyle{definition}
\newtheorem{theorem}{Theorem}[section]

\newtheorem{proposition}[theorem]{Proposition}
\newtheorem{definition}[theorem]{Definition}

\title{Intersecting hypersurfaces in AdS and Lovelock gravity}
\author{Elias Gravanis$^1$\footnote{E-mail:
eliasgravanis@netscape.net} and Steven Willison$^{1,2}$ \footnote{E-mail:
steve-at-cecs.cl}\\\\
{\it $^1$Department of Physics, Kings College, Strand, London WC2R
2LS, U.K.}
\\
{\it $^2$Centro de Estudios Cient\'{\i}ficos (CECS), Casilla 1469,
Valdivia, Chile.} }
\date{}

\begin{document}

\maketitle

\begin{abstract}

Colliding and intersecting hypersurfaces filled with matter
(membranes) are studied in the Lovelock higher order curvature
theory of gravity. Lovelock terms couple hypersurfaces of
different dimensionalities, extending the range of possible
intersection configurations. We restrict the study to constant
curvature membranes in constant curvature AdS and dS background
and consider their general intersections. This illustrates some
key features which make the theory different to the Einstein
gravity. Higher co-dimension membranes may lie at the intersection
of co-dimension 1 hypersurfaces in Lovelock gravity; the
hypersurfaces are located at the discontinuities of the first
derivative of the metric, and they need not carry matter.

The example of colliding membranes shows that general solutions
can only be supported by (spacelike) matter at the collision
surface, thus naturally conflicting with the dominant energy
condition (DEC). The imposition of the DEC gives selection rules
on the types of collision allowed.

When the hypersurfaces don't carry matter, one gets a soliton-like
configuration. Then, at the intersection one has a co-dimension 2
or higher membrane standing alone in AdS-vacuum spacetime \emph{
without conical singularities.}

Another result is that if the number of intersecting hypersurfaces
goes to infinity the limiting spacetime is free of curvature
singularities if the intersection is put at the boundary of each
AdS bulk.

\end{abstract}


\section{Introduction}

Lately, a strange idea has become popular in cosmology. It has been
suggested~\cite{Randall} that we live on a $(3+1)$-dimensional
membrane, called a brane world, living in a higher dimensional
space-time. Many general relativity models have been invented to
describe the gravitational behavior of such a brane-world. Although
there is a clear conceptual link with string theory, i.e. the extra
dimensions and the existence of membranes with matter and gauge
fields confined to their world-sheets, it is also clear that this is
a highly speculative idea.


This idea motivates a  general study of hypersurfaces in
$d$-dimensional curved spacetime. Co-dimension 1 hypersurfaces are
understood as co-dimension 1 sub-manifolds which are the locus of
the discontinuities of the first derivative of the metric. To draw
specific conclusions we need a theory of gravity, determining the
metric of the $d$-dimensional spacetime locally. Lovelock gravity is
a natural choice in $d$ dimensions in place of Einstein gravity in
four dimensions; it is the only theory (action functional) for the
metric which gives second order field equations when torsion is
zero, that is, when the covariant derivative is given by the usual
formula~\cite{Zumino-86}\cite{Lovelock-71}. One can get a relation
between the discontinuity of the first derivative of the metric to
the energy tensor of matter on the hypersurface. Hypersurfaces of
any co-dimensionality which (potentially) carry matter will be
called membranes.

The more complicated, compared to Einstein's theory, structure of
derivatives in Lovelock gravity, makes possible to have membranes of
co-dimensionality higher than one, via intersections of co-dimension
1 hypersurfaces, without any spacetime singularities. Put slightly
differently, high co-dimension membranes can be embedded in
spacetime without causing conical or more pathological curvature
singularities if they are embedded at the intersection of
co-dimension 1 hypersurfaces. In fact, in $d$ dimensions there exist
membranes of co-dimensionality up to the integer part of
$\frac{d-1}{2}$, such that the metric is everywhere continuous, its
first derivative has (bounded) discontinuities at the hypersurfaces
and spacetime is everywhere, and especially at the membranes, a
manifold~\cite{Gravanis-03}\cite{Gravanis-04}.

The higher dimensional gravity theory of Lovelock~\cite{Lovelock-71}
is an interesting generalization of general relativity. In $d\geq 5$
the Einstein-Hilbert is not the most general Lagrangian that
produces second order field equations and it was extended by
Lovelock to a more general theory with this property. The latter
gives the theory  familiar features, in accordance with our
experience from classical mechanics and field theory. It allows for
a Hamiltonian formulation~\cite{Teitelboim-87} and the possibility
of a well-posed initial value problem~\cite{Choquet-Bruhat-88}. The
Lagrangian which possesses this property was found by
Lovelock~\cite{Lovelock-71} and it is a linear combination of terms
corresponding to the Euler densities in all lower even
dimensions~\cite{Zumino-86}.
\begin{gather} {\cal L} =
\sum_{n=0}^{[(d-1)/2]} \frac{1}{2^n}\beta_n \delta^{\mu_1...
\mu_{2n}}_{\nu_1...\nu_{2n}}
R^{\nu_1\nu_2}_{\mu_1\mu_2}\cdot\cdot\cdot
R^{\nu_{2n-1}\nu_{2n}}_{\mu_{2n-1}\mu_{2n}}
\sqrt{g}\,d^dx.\label{Lovelockscalar}
\end{gather}
where $[x]$ is the integer part $x$. The generalization of the
Einstein tensor is the Lovelock tensor:
\begin{gather}
H^{\mu}_{\nu} = -\sum_{n=0}^{[(d-1)/2]}\frac{1}{2^{n+1}}
\beta_n\delta^{\mu\mu_1... \mu_{2n}}_{\nu\nu_1...\nu_{2n}}
R^{\nu_1\nu_2}_{\mu_1\mu_2}\cdot\cdot\cdot
R^{\nu_{2n-1}\nu_{2n}}_{\mu_{2n-1}\mu_{2n}}
\end{gather}

The delta is the generalized totally anti-symmetrized Kronecker
delta. It is the determinant of a matrix with elements $\delta^M_N$,
\begin{gather}
\delta^{\mu_1... \mu_{p}}_{\nu_1...\nu_{p}} =\det \left(
       \begin{array}{cccc}
        \delta^{\mu_1}_{\nu_1} & \delta^{\mu_2}_{\nu_1}
        &\ldots & \delta^{\mu_p}_{\nu_1} \\
        \delta^{\mu_1}_{\nu_2} & \delta^{\mu_2}_{\nu_2}
        & \ldots & \delta^{\mu_p}_{\nu_2} \\
        \vdots  & \vdots  & \ddots & \vdots \\
        \delta^{\mu_1}_{\nu_p} & \delta^{\mu_2}_{\nu_p}
        & \ldots & \delta^{\mu_p}_{\nu_p}
       \end{array}
     \right)
=p!\delta^{\mu_1}_{[\nu_1}\cdots \delta^{\mu_{p}}_{\nu_{p}]}.
\end{gather}

The Lovelock theories have been studied extensively. Higher
dimensional black hole solutions have been
found~\cite{Boulware-85}\cite{dcblackholes}\cite{Crisostomo}. This
has shed some interesting light on questions of black hole entropy.
Some cosmological metrics have been studied~\cite{Madore-86}.

The $n_{\text{max}} = 2$ Lovelock theory, which we call the
Gauss-Bonnet theory, has a special physical significance. This is
because the $n=2$ term is the only quadratic term which has a ghost
free perturbation theory about flat space-time. It has been
conjectured that the Gauss-Bonnet term is the leading order, purely
geometric, correction to the effective action of an underlying
unitary fundamental theory~\cite{Zwiebach-85}. In particular, the
Lovelock contributions, motivated by string theory, have played a
role in brane-world cosmology~\cite{Mavromatos-00,Davis-02}.

It was Zumino~\cite{Zumino-86} who formulated the theory in the way
we prefer, as an elegant way to prove suggestions by Zwiebach
related to low energy string theory~\cite{Zwiebach-85}. We use the
vielbein formulation: $E^a$ is the vielbein frame, $\omega^{ab}$ the
spin connection and $\Omega^{ab}$ is the curvature two-form.
\begin{gather*}
\Omega^{ab} = \frac{1}{2}R^{ab}_{\ \ cd}E^c\wedge E^d,
\\ R^{ab}_{\ \ cd} = E^a_\mu E^b_\nu R^{\mu\nu}_{\ \ \kappa\lambda}
E^\kappa_c E^\lambda_d.
\end{gather*}
In this language, the Lovelock Lagrangian is:
\begin{gather}\label{genLove}
{\cal L}= \sum_{n=0}^{[(d-1)/2]} \beta_n\Omega^{a_1
a_2}\wedge\cdots\Omega^{a_{2n-1}a_{2n}} \wedge e_{a_1\cdots a_{2n}}
\end{gather}
where
\begin{gather}\label{volform}
e_{a_1\cdots a_p} = \frac{1}{(d-p)!}\, \epsilon_{a_1\dots a_d}
E^{a_{p+1}}\wedge \cdots \wedge E^{a_{d}}
\end{gather}
and we have defined the totally anti-symmetric tensor such that
$\epsilon_{(1)\dots (d)} = 1$. The Latin letters from the beginning
of the alphabet are used for the local Lorentz indices
($d$-dimensional). Greek letters from the middle of the alphabet are
used for space-time co-ordinate indices ($d$-dimensional).

In the Lovelock theory, singular hypersurfaces of co-dimension 1 can
be meaningfully defined in terms of distributions
~\cite{Davis-02,Maeda-03}, due to the property of quasi-linearity in
second derivatives~\cite{Deruelle-03}. Brane-worlds of co-dimension
$1$ have thus been the most well studied and understood. They can
also be formulated by means of boundary terms in the action. The
correct boundary term is most elegantly derived by a dimensional
continuation of the Gauss-Bonnet theorem for a manifold with
boundary~\cite{Myers-87}. This latter approach is the one we have
adopted.

The possibility of colliding shells or branes of matter has been
studied in the context of GR~\cite{Dray-86}. In Lovelock gravity,
there has been some study of intersecting brane-worlds, see
e.g.~\cite{Kim-01a} and more
recently~\cite{Lee-04}\cite{Navarro-04}, but so far, there has been
no study of collisions in this context except our comments in our
recent work~\cite{Gravanis-03,Gravanis-04}. In that work, we
restricted the smoothness of the metric so that there were well
defined ortho-normal vectors at the intersection/collision. The most
striking fact, physically, about intersections or collisions is that
they could carry their own singular stress-energy tensor. This is a
phenomenon that does not occur in the Einstein theory. That
difference and related properties of the Gauss-Bonnet term was used
in~\cite{Kim-01a}, to address the cosmological constant problem and
formulate higher co-dimension brane worlds via intersections,
continuing on previous work in GR context, see
e.g.~\cite{Arkani-Hamed:1999hk}. Our aim here will be to discuss
general properties of the intersections and collisions of
hypersurfaces in Lovelock gravity, mostly via the example of AdS and
dS background, which may be useful also outside the brane-world
context.

In the Einstein theory, singular matter can only be accommodated at
an intersection of co-dimension 2 if there is a conical singularity,
with a deficit angle. Then, it is impossible to define two
ortho-normal vectors normal to the intersection. Although Lovelock
gravity with a conical singularity can be described in terms of
distributions~\cite{Fursaev-95}\cite{Bostock-03}\cite{Charmousis:2005ey},
there is a certain ambiguity about the solutions- in general, we
would not expect the thin brane to be the unique limit of a thick
brane solution~\cite{Geroch-87,Garfinkle-99}. We shall not consider
this kind of singularity in the present work.


There is though an interesting possibility. We consider
intersections of hypersurfaces, non-null as well as null, which
carry zero energy tensor. At their intersections there appear higher
co-dimension membranes. For non-null co-dimension 1 hypersurfaces we
have an intersection of soliton-like configurations, pure
(cosmological constant-) vacuum gravitational field self-supported
and with a non-zero jump in the extrinsic curvature; for null
hypersurfaces we have intersection/collision of gravitational shock
waves. In both cases one has at their intersections membranes of
co-dimension $\geq 2$ surrounded by pure AdS background on a
non-singular spacetime. This is a phenomenon not possible in
Einstein gravity~\cite{Geroch-87}.

Previous works on related problems in the brane-world were in the
context of: Einstein gravity e.g.~\cite{Arkani-Hamed:1999hk};  in
supergravity, where intersection rules for branes carrying form
field charges were derived in
refs.~\cite{Argurio:1997gt}\cite{Ohta:1997gw}\cite{Ohta:2003uw}; and
in various formulations when Gauss-Bonnet or higher Euler densities
are included,
e.g.~\cite{Iglesias:2000iz}\cite{Kim-01a}\cite{Lee-04}\cite{Navarro-04}\cite{Bostock-03}.
An important difference in this work and our previous
ones~\cite{Gravanis-03}\cite{Gravanis-04} is that one may have a
high co-dimension membrane without the cost of making spacetime
singular~\cite{Iglesias:2000iz}\cite{Bostock-03}\cite{Navarro-04}.
Our primary intention really is to point out properties of Lovelock
gravity which are interesting on their own, but our results may be
useful to other endeavors.

In sections \ref{intersections}-\ref{non_simp_Section} we present
the example of intersecting hypersurfaces in an anti-de Sitter
background. In section \ref{DEC_Section} we discuss colliding
hypersurfaces in de Sitter background and point out the spontaneous
dominant energy condition violation in collisions. In section
\ref{dimension_Section}, we discuss the dimensionalities of the
intersection in relation to a 4-dimensional universe. In section
\ref{high} we discuss higher co-dimension membranes using
intersections of solitonic configurations and shock waves.

\subsection{The intersection junction conditions}

For our purposes, hyper-surfaces are $(d-1)$-dimensional surfaces
which divide the space-time up into $d$-dimensional bulk regions. We
shall assume that they are space-time like (i.e. with space-like
normal vector). If there is a non-zero singular component to the
stress-energy tensor with it's support on the hyper-surface, we
shall also call it a brane.

The mathematics of the intersections becomes simple if we consider a
kind of minimal intersection, which involves the minimum number of
hypersurfaces needed to build the intersection of a given
co-dimensionality (dimensionality of its normal space). Put
differently, in such an intersection any bulk region has (a
co-dimension 1) common boundary with \emph{any other} bulk region.
Not without a reason we call them simplicial intersections: if
abstractly we assign a point to every bulk region in which the
connection is continuous, then a co-dimension $p$ intersection
corresponds to a $p$-dimensional simplex, that is, the
$p$-dimensional polyhedron with the minimum number of vertices. This
abstraction turns into a practical method of calculating the
Lagrangian densities integrated over the intersections
~\cite{Gravanis-03,Gravanis-04}.

One of the simplifications related to the simplicial intersection,
is that if we label the bulk regions with $i$ (and designate
$\{i\}$) then the co-dimension $p$ intersection can be labeled by an
anti-symmetric symbol involving the labels of the $p+1$ bulk regions
meeting there; the simplest example are the co-dimension 1
hyper-surfaces designated in general as $\{i_0i_1\}=-\{i_1i_0\}$. So
we introduce the following ~\cite{Gravanis-03,Gravanis-04}
\begin{definition}(simplicial intersection)
Let $\{i\}$ be a bulk region. $\{i_0\dots i_p\}$ is a simplicial
intersection where bulk regions $i_0, \dots, i_p$ meet, if it is a
$(d-p)$-dimensional submanifold. The connections in the bulk regions
are $\omega_0,\dots,\omega_p$ respectively. $\{i_0\dots i_p\}$ is a
part of the boundary of the $(p-1)$-intersection $\{i_0\dots
i_{p-1}\}$. The orientation is $\partial \{i_0...i_{p-1}\} =
+\{i_0...i_p\}+\cdots$. Swapping any pair of indices reverses the
orientation.
\end{definition}
\noindent Note that intersections may be space-like, time-like or
null (or vary between them).

There are junction conditions relating the singular stress-energy to
the geometry~\cite{Gravanis-03,Gravanis-04}. Let $\omega_i$ be the
connection in region $\{i\}$. At a hypersurface $\{ij\}$ there can
be a discontinuity $\omega_i \neq \omega_j$. The junction conditions
at a $p$-intersection are obtained from the intersection Lagrangian:
\begin{gather} \label{A}
\sum_{n=1}^{[(d-1)/2]} \beta_n {\cal L}^n_{(p)},
\\\nonumber
{\cal L}^n_{(p)}(E,\omega_0,..,\omega_{p}) = A_p
\int_{s_{0...p}}\hspace{-.2in} d^pt\,
(\omega_1-\omega_0)^{a_1b_1}\cdots
(\omega_{p}-\omega_{0})^{a_pb_p}\,
\Omega(t)^{a_{p+1}b_{p+1}...a_nb_n} e_{a_1...b_n},
\\\nonumber
A_p = (-1)^{p(p-1)/2} \frac{n!}{(n-p)!}.
\end{gather}
$\Omega(t)$ is the curvature of the interpolating connection
$\omega(t)$:
\begin{equation}\label{omegatOmegat}
\omega(t):=\sum_{i=0}^p t^i \omega_i \quad , \quad
\Omega(t)=d\omega(t)+\omega(t) \wedge \omega(t)
\end{equation}
The $\Omega(t)^{a_{p+1}\dots b_n}$ is short for the $(n-p)$-fold
product: $\Omega(t)\wedge\cdots\wedge \Omega(t)$. The integral is
over the $p$-dimensional simplex
\begin{equation}
s_{0\dots p}=\{ t \in \mathbf{R}^{p+1} \,| \sum_{i=0}^p t_i =1, \
{\rm all} \ t_i \geq 0\}.
\end{equation}

The junction conditions come from explicit Euler variation w.r.t. to
the vielbein: $ \delta_{E^c} {\cal L}_{(p)} = -2 (T_{(d,d-p)})^b_c
{\tilde e}_b$, where $T_{(d,d-p)}$ is the part of the singular
stress-energy tensor with support on the intersection. The factor of
$-2$ is explained in the Appendix. So the junction conditions
\footnote{
In Einstein theory the only junction condition is that of the
hypersurface, where if the energy tensor carried by it vanishes then
so does the discontinuity of the connection, in the non-null case.
For a general intersection in Lovelock gravity the energy tensor may
vanish without implying vanishing of the connection discontinuities.
Even for the single hypersurface case, ${\cal E}_{(1)}=\sum_n
\beta_n {\cal E}^n_{(1)}=0$ does not imply that the connection
becomes continuous. Simple solutions where this happens can easy be
found and such spacetimes have been called solitons
~\cite{Iglesias:2000iz}. If such a hypersurface is spacelike, there
is a breakdown of causality. Another important case of vanishing
energy tensor is that of the null hypersurface, that is, of the
shock wave. Shock waves exist in GR ~\cite{Penrose} as well as in
higher order Lovelock theory.} can be written as:
\begin{gather}
(T_{(d,d-p)})^b_c {\tilde e}_b = \frac{1}{2}
(-1)^{(p-1)(p-2)/2}\sum_{n=p}^{[(d-1)/2]}\frac{n!}{(n-p)!} ({\cal
E}^n_{(p)})_c, \label{Thejunction}
\\
({\cal E}^n_{(p)})_c \equiv \int_{s_{1...p+1}} d^pt
(\omega_2-\omega_1)^{a_1b_1}\cdots
(\omega_{p+1}-\omega_{1})^{a_pb_p}
\Omega(t)^{a_{p+1}b_{p+1}...a_nb_n} e_{a_1...b_nc}.
\end{gather}
${\tilde e}$ is the natural volume element on the intersection. We
note that ${\cal E}^n_{(p)}$ is zero if $p>n$.

Also there is another implicit junction condition: there is a well
defined (pseudo) ortho-normal frame everywhere. If this condition is
not obeyed, then the above formula is not valid. In the case of a
hyper-surface junction condition, it is equivalent to a well-defined
induced geometry on the hypersurface. For higher co-dimension
intersections it is a quite stringent condition. For example, for a
co-dimension 2 intersection, there can be no deficit angle.


\section{Intersections in AdS and Gauss-Bonnet term}\label{intersections}

We have seen that there is a possibility to localize matter on an
intersection in the Gauss-Bonnet theory. We now proceed to a
specific example.

\subsection{The bulk vacuum solution}

We shall take the simplest kind of bulk solution. Each bulk region
is a constant curvature region of space-time. Such a space-time
satisfies $R_{\mu\nu\rho\sigma} = \frac{1}{d(d-1)}R(g_{\mu\rho}
g_{\nu\sigma} -g_{\mu\sigma} g_{\nu\rho})$, $R$ being a
constant~\cite{Hawking-73}. There are three possibilities:
\\
\begin{tabular}{ll}
i) &  de Sitter space ($R>0$),\\
ii) & anti-de Sitter space ($R<0$),\\
iii) & flat space ($R=0$).\\
\end{tabular}

In the Einstein theory, constant curvature empty space will be one
of the above three, depending on whether the cosmological constant
is positive, negative or zero. In the higher order Lovelock theory,
it is possible that more than one type of constant curvature
space-time will satisfy the vacuum field equations. The different
possibilities arise because the field equations are polynomial in
the curvature. For a constant curvature, this just reduces to a
polynomial equation in the curvature scalar.

A more general space-time would be made up of regions of less
symmetric vacuum space time. We will not attempt this here, but
leave it as a project for the future.

We take the example of anti-de Sitter (AdS) bulk space-time,
motivated by: 1) the Randall-Sundrum idea of the non-factorizable
metric ~\cite{Randall} which allows gravitons to be approximately
localized in a large extra dimension; 2) the special role of AdS
space in recent advances ~\cite{Maldacena-97}; 3) The simplicity of
the problem from a mathematical point of view. Anti-de-Sitter space
has constant negative curvature:
\begin{gather}\label{AdScurve}
\Omega^{ab} = -\frac{1}{l^2}E^a\wedge E^b,
\end{gather}
The constant $l$ has dimensions of length. It is easy to check that
if we write:
\begin{gather}\label{foundconnection}
\omega^{ab} = \frac{1}{l}(u^a E^{b} -u^b E^{a}).
\end{gather}
where $u^a$ is a constant vector, we have, assuming zero torsion
$dE^a = -\omega^a_{\ b}\wedge E^b$,
\begin{gather}
\Omega^{ab} = -\frac{u^2}{l^2} E^a\wedge E^b.
\end{gather}
Above $u^2 = \eta_{ab}u^au^b$. For an AdS solution, we take $u$ to
be space-like $u^2=+1$. The opposite sign choice gives dS spacetime.
AdS, dS or flat space is a vacuum solution of the general Lovelock
theory (\ref{genLove}) provided that the following relation is
satisfied:
\begin{gather} \label{bulkeq}
\sum_{n=0}^{[(d-1)/2]}\frac{(-u^2)^n (d-1)!}{(d-1-2n)!} \frac{
\beta_n}{l^{2n}} =0.
\end{gather}

Now let us write the solution in terms of co-ordinates. We will
write the AdS metric in conformally flat form. Define $u\cdot x
\equiv \eta_{\mu\nu}u^{(\mu)}x^\nu$.
\begin{gather}\label{metric}
ds^2 = \frac{1}{((u\cdot x)/l +C)^2} \ \eta_{\mu\nu}dx^\mu
dx^\nu,\qquad (u\cdot x)/l +C > 0.
\end{gather}
With $C$ an arbitrary constant. Contact between (\ref{metric}) and
(\ref{foundconnection}) is made by the choice for the vielbein:
\begin{gather*}
E^{a} = \frac{1}{(u\cdot x)/l +C}\ \delta^{a}_{(\mu)}\ dx^\mu.
\end{gather*}

We will only be interested in the vicinity of the intersection and
will not worry here about the global details of joining together
regions of AdS.

\subsection{Three-way intersection}\label{3way}

We will consider the simplest 3-way vertex. There is a plane covered
by co-ordinates $(x,y)\equiv (x^{d-2},x^{d-1})$. It will also be
convenient to use cylindrical co-ordinates:  $x = \rho
\cos{\theta}$, $y=\rho\sin{\theta}$. There are $3$ bulk regions, $i
= 1,2,3$, broken up by $3$ hypersurfaces, $\{ij\}$, at $\theta =$
const. The hypersurfaces meet at the intersection, $\{123\}$, at
$\rho=0$. The space-time is divided into regions:
\\
Region 1: $0 < \theta < \theta_1$,
\\
Region 2: $\theta_1 < \theta < \theta_2$, \\Region 3:
$\theta_2<\theta<\theta_3$, \\with the identification $\theta_3
\equiv 0$. One can have a conical singularity at the intersection
with deficit angle $2\pi - \theta_3$ but we will not do so for
reasons we will mention. So we take $\theta_3 =2\pi$ (fig. 1).
\begin{figure}\label{nodeficit}
\begin{centering}
\begin{picture}(0,0)%
\includegraphics{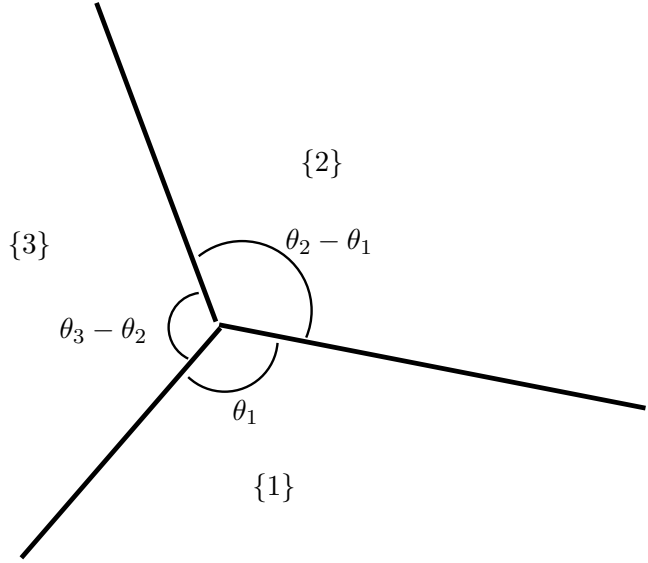}%
\end{picture}%
\setlength{\unitlength}{3947sp}%
\begingroup\makeatletter\ifx\SetFigFont\undefined%
\gdef\SetFigFont#1#2#3#4#5{%
  \reset@font\fontsize{#1}{#2pt}%
  \fontfamily{#3}\fontseries{#4}\fontshape{#5}%
  \selectfont}%
\fi\endgroup%
\begin{picture}(4039,3554)(2489,-3401)
\put(4232,-1429){\makebox(0,0)[lb]{\smash{{\SetFigFont{11}{13.2}{\rmdefault}{\mddefault}{\updefault}$\theta_2-\theta_1$}}}}
\put(3901,-2499){\makebox(0,0)[lb]{\smash{{\SetFigFont{11}{13.2}{\rmdefault}{\mddefault}{\updefault}$\theta_1$}}}}
\put(2814,-1986){\makebox(0,0)[lb]{\smash{{\SetFigFont{11}{13.2}{\rmdefault}{\mddefault}{\updefault}$\theta_3-\theta_2$}}}}
\put(2489,-1449){\makebox(0,0)[lb]{\smash{{\SetFigFont{11}{13.2}{\rmdefault}{\mddefault}{\updefault}$\{3\}$}}}}
\put(4026,-2974){\makebox(0,0)[lb]{\smash{{\SetFigFont{11}{13.2}{\rmdefault}{\mddefault}{\updefault}$\{1\}$}}}}
\put(4326,-936){\makebox(0,0)[lb]{\smash{{\SetFigFont{11}{13.2}{\rmdefault}{\mddefault}{\updefault}$\{2\}$}}}}
\end{picture}%
\caption{{\small The intersection of three hypersurfaces. Each
bulk region, denoted $\{i\}$, is a piece of constant curvature
space-time. If there is no deficit angle we have $\theta_3 =
2\pi$.} }
\end{centering}
\end{figure}
In each region $i$ let $u_i = (0,...,0,\cos{\phi_i},\sin{\phi_i})$
such that $u_i\cdot x = \rho \cos{(\theta -\phi_i)}$. The metric in
each region takes the following form:
\begin{gather}
ds_i^2 = \frac{1}{(\rho\cos(\theta-\phi_i)/l +1)^2} \
\eta_{\mu\nu}dx^\mu dx^\nu.
\end{gather}
We have chosen $C=1$ here for convenience. We insist that the metric
is continuous, so the factor $(u\cdot x)/l +1$ should be continuous
across the walls:
\begin{align}\label{continuity conditions in angles}
\cos(\theta_1-\phi_1) =& \cos(\theta_1-\phi_{2}),
\\\cos(\theta_2 - \phi_{2})=&
\cos(\theta_{2}-\phi_{3}),
\\ \cos(\phi_{3})=& \cos(\phi_{1}).
\end{align}
There is the trivial solution $\phi_i = \phi_{i+1}$, which is smooth
across the hypersurface. If we are to have any matter on the
hypersurfaces (a brane) we must choose the non-smooth solutions:
\begin{gather}
\phi_1= -\phi_3=\theta_1-\theta_2\nonumber\\
\phi_2=\theta_1 +\theta_2\label{angles}
\end{gather}
This allows for $u$ to be different in each region. The
spin-connection (\ref{foundconnection}) is not single-valued at the
walls.

At the intersection, we need more than just continuity of the
metric. We must have a well defined ortho-normal basis.
\begin{gather*}
E^{a}_\mu E^{b}_\nu g^{\mu\nu} =\eta^{ab}
\end{gather*}
everywhere, including at $\rho =0$. Now, since the metric is
conformally flat, the angle between two vectors is:
\begin{gather*}
v \angle w = \frac{v^\mu w^\nu g_{\mu\nu}} {\left(v^{\mu}v^\nu
g_{\mu\nu}\right)^{1/2} \left( w^{\mu}w^\nu g_{\mu\nu}\right)^{1/2}
} = \frac{v^\mu w^\nu \eta_{\mu\nu}} {\left(v^{\mu}v^\nu
\eta_{\mu\nu}\right)^{1/2} \left( w^{\mu}w^\nu
\eta_{\mu\nu}\right)^{1/2} }
\end{gather*}
This is well known- a conformal transformation preserves angle. So
theta is indeed a measure of the angle between vectors. At $\rho
=0$, we have $E^a = (\dots,\cos\theta_a,\sin\theta_a)$, but with the
identification $\theta \equiv \theta + \theta_3$. Since $\eta^{ab} =
E^a\angle E^b = \cos(\theta_b-\theta_a)$, for a well defined
ortho-normal frame we require $\cos(\theta_b-\theta_a+r\theta_3) =
1\ or\ 0$ for arbitrary integer $r$ , so we should set $\theta_3 =
2\pi$. Thus we insist upon having no deficit angle at the
intersection.

\subsection{The junction conditions}
Recall, the junction condition at each $p$-intersection is
(\ref{Thejunction}). In the next section we shall evaluate the
general form of ${\cal E}^n$ for the intersections in AdS. Here we
shall stick to the co-dimension $2$ intersection in the
Einstein-Gauss-Bonnet theory. Furthermore, we shall not calculate
the energy-momentum tensor on the branes but proceed to find what is
at the intersection.

There is no contribution from the Einstein term: ${\cal
E}^1_{(2)}=0$. Only the Gauss-Bonnet contributes. ${\cal E}^2_{(2)}$
is:
\begin{gather*}
({\cal E}^2_{(2)})_f = (\omega_2-\omega_1)^{ab}\wedge
(\omega_3-\omega_1)^{cd}\wedge {e}_{abcdf} \int_{s_{123}} d^2t.
\end{gather*}
The volume of the $2$-simplex is $1/2$.
\begin{align*}
({\cal E}^2_{(2)})_f = & \frac{1}{2}(\omega_2-\omega_1)^{ab}\wedge
(\omega_3-\omega_1)^{cd} \wedge {e}_{fabcd}\\
= & -2 (u_2-u_1)^{a}(u_3-u_1)^{b}
E^c\wedge E^{d} \wedge {e}_{fabcd}\\
= & -2(d-4)(d-3)\left[(u_2-u_1)^{(1)}(u_3-u_1)^{(2)}
-(u_2-u_1)^{(2)}(u_3-u_1)^{(1)}\right] {e}_{f(1)(2)}.
\end{align*}
The factor in square brackets is
\begin{align*}
&(\cos\phi_2-\cos\phi_1)(\sin\phi_3-\sin\phi_1)
-(\cos\phi_3-\cos\phi_1)(\sin\phi_2-\sin\phi_1)
\\&=  \sin(\phi_3-\phi_2) +\sin(\phi_2-\phi_1)
+\sin(\phi_1-\phi_3)\\
&= \sin(2\theta_2) +\sin(2\theta_1-2\theta_2)-\sin(2\theta_1).
\end{align*}
Note ${e}_{(1)(2)}=\tilde{e}$ is the natural volume element on the
intersection. Putting this into (\ref{Thejunction}), we get the
following result:
\begin{proposition}
The junction condition for the intersection is:
\begin{gather}\label{examplelast}
({T}_{123})^a_b = -2(d-4)(d-3)\beta_2 [\sin(2\theta_2)
+\sin(2\theta_1-2\theta_2)-\sin(2\theta_1)]\delta^a_b
\end{gather}
The singular matter on the intersection is a $(d-2)$-dimensional
cosmological constant or tension.\footnote{ If the energy tensor
$T^a_b$ on a hypersurface takes the form $- V \delta^a_b$, we will
call the constant $V$ the cosmological constant or tension of the
membrane, which in some cases might be negative where it amounts to
pressure. In this paper we are mainly interested in whether the
value of the energy tensor is zero or not and we will be writing the
tensor as $T^a_b=\Lambda \delta^a_b$ so one should bear in mind that
$V=-\Lambda$.}
\end{proposition}

Using the double angle formulas, we can prove that this tension
vanishes in $d\geq 5$ if and only if:
either $\cos(2\theta_1)=1$, $\cos(2\theta_2)=1$ or
$\cos(2\theta_1)=\cos(2\theta_2)$ and
$\sin(2\theta_1)=\sin(2\theta_2)$. These solutions are not really
intersections at all:
\\ i) $\theta_1 =\theta_2 \Rightarrow
\phi_1=\phi_3 =0$, region 2 is shrunk to zero; $\theta_1 =0
\Rightarrow \phi_2=\phi_3$, region 1 is shrunk to zero; or $\theta_2
=2\pi \Rightarrow \phi_2=\phi_1$, region 3 is shrunk to zero. In
these
cases there is just a smooth AdS bulk.\\
ii) $\theta_1=\pi \Rightarrow \phi_2=\phi_3$, $\theta_2=\pi
\Rightarrow \phi_2=\phi_1$ or $\theta_1 - \theta_2 =\pi\Rightarrow
\phi_3=\phi_1=\pi$ In these cases there is just a single
hypersurface.



\section{ Higher co-dimension intersections in
AdS}\label{higherintersections}

So far we have dealt with co-dimension $2$ intersections. We now
proceed to look at the higher co-dimension simplicial intersection
in AdS background. There are $p+1$ bulk regions, $\{i\}$, separated
by $p+1$ hypersurfaces, $\{ij\}$, intersecting at the simplicial
intersection $\{i_0 \dots i_p\}$.
The metric in each bulk region is, c.f. (\ref{metric}):
\begin{gather}
ds^2 = \frac{1}{(u_i\cdot x+1)^2}\left(\delta_{\alpha\beta}dx^\alpha
dx^{\beta} + dx^2_{(d-p)}\right).
\end{gather}
The branes intersect at $x^\alpha=(0,\dots,0)$. Each brane is
parameterized generally by $f(x^{\alpha})=0$, and is assumed to be
maximally symmetric in the other $d-p$ dimensions.

The continuity of the metric at each hypersurface $\{ij\}$: $(u_i -
u_j)\cdot x = 0$, implies that $(u_i - u_j)$ is proportional to the
normal vector to $\{ij\}$. Each AdS region is characterized by a
unit spacelike vector $u^a_i, i=0,..,p$ and the same AdS scale $l$
which we set to 1. Define
\begin{equation}\label{uij}
u_{ij}=u_i-u_j
\end{equation}
and
\begin{align}
u(t)= & \sum_{i=0}^p t^i u_i
\end{align}
with $\sum_{i=0}^p t^i=1$, and
\begin{equation}\label{defineN}
R(t)^a_b=u(t)^2 \delta^a_b+N^a_b \quad ,  \quad N^{ab}=
\sum_{i=0}^p\sum_{j=0}^p t^i t^j u^a_{ij} u^b_{ij}
\end{equation}
\begin{proposition}
The curvature of the interpolating connection is
\begin{gather}\label{adsinterpcurv}
\Omega^{ab}(t)=-R(t)^{c[a} E_c \wedge E^{b]},
\end{gather}
where the symmetric matrix $R(t)$ is defined in (\ref{defineN}).
\end{proposition}
\emph{Note} : We will {\it not} need to take the $u_i$ all of the
same causal nature in this proof; each $u_i$ may be time- or
space-like, or null.

\emph{Proof} : By (\ref{foundconnection}) the connection on each
region $i$ is assumed to be given by
\begin{equation}
\omega^{ab}_i=u_i^a E_i^b-u_i^b E_i^a
\end{equation}
The curvature of the interpolating connection
$\omega(t)=\sum_{i=0}^p t^i \omega_i$ then is
\begin{eqnarray}
&& \Omega(t)^{ab}=d \omega(t)^{ab}+\omega(t)^a_{\ c} \wedge
\omega(t)^{cb}=-u(t)^2 E^a \wedge E^b+ \\\nonumber &&+ \sum_{i=0}^p
t^i u^a_i u_{ic} E^b\wedge E_c + \sum_{ij} u^a_i u_{jc} E^c \wedge
E^b - (a \leftrightarrow b) \nonumber
\end{eqnarray}
where we have used zero torsion and metric continuity to calculate
\begin{equation}
d\omega(t)^{ab}=\sum_i t^i d \omega^{ab}_i= \sum_i t^i (u^a_i
dE_i^b-u^b_i dE_i^a)=\sum_i t^i (-u^a_i \omega^b_{i c}\wedge
E^c+u^b_i \omega^a_{ic} \wedge E^c)
\end{equation}
where we drop the region index from the frame $E$ after the
derivative is taken as the metric itself is continuous, all $E_i$
agree at the hypersurface, only its derivative jumps.

Now, by using $\sum_{i=0}^p t^i=1$, we have
\begin{equation}
\sum_{ij} t^i t^j (u_i-u_j)^a (u_i-u_j)_c=2 \sum_i t^i u^a_i
u_{ic}-2\sum_{ij} t^i t^j u^a_i u_{jc}
\end{equation}
so by (\ref{defineN}) and (\ref{uij}) we get
(\ref{adsinterpcurv}).$\Box$

The intersection junction conditions are (\ref{Thejunction}) with:
\begin{gather}
{\cal E}^n_{(p)} = \ 2^{p} \ (-1)^{n-p} \
  \int_{s_{01..p}} d^pt \;
u^{a_1}_{10}..u^{a_p}_{p0} \: E^{b_1..b_p} \:
R(t)^{a_{p+1}}_{c_{p+1}}..R(t)^{a_n}_{c_n} \:
E^{c_{p+1}b_{p+1}..c_nb_n} \:
  e_{a_1...b_nc}.
\end{gather}
Note first that by $u_{ij}=u_{i0}-u_{j0}$ all terms involving
$N^a_b$ defined by (\ref{defineN}) and (\ref{uij}) drop out in the
previous equation by the presence of the factors $u_{10}..u_{p0}$
(involving all vectors which span the normal space) and the
anti-symmetry of the volume form $e_{a_1..b_nc}$. Note that this
also true if some $u_{i0}$'s are null. Now applying the identity
\begin{gather}
E^{c_1...c_n} \wedge e_{d_1...d_m}=\frac{m!}{(m-n)!}
\delta^{c_1}_{[d_{m-n+1}}...\delta^{c_n}_{d_m} \:
e_{d_1...d_{m-n}]},
\end{gather}
we can then write
\begin{multline} \label{lastbefore}
{\cal E}^n_{(p)} =
  (-1)^{n-p}(-1)^{p(p-1)/2} \ 2^p  \
  \int_{s_{01..p}} d^pt \;
u^{a_1}_{10}..u^{a_p}_{p0} \: \: (u(t)^2)^{n-p} \:\\ \times
\frac{(2n+1)!}{(p+1)!} \:
\delta^{b_1}_{[b_1}\dots\delta^{b_p}_{b_p}\delta^{a_{p+1}}_{a_{p+1}}\:
\delta^{b_{p+1}}_{b_{p+1}}\dots\delta^{a_n}_{a_n}\:\delta^{b_n}_{b_n}
\: e_{ca_1\dots a_p]}
\end{multline}
The factor of $(-1)^{p(p-1)/2}$ comes from the rearrangement of the
indices. The quantity after the symbol $\times$ equals
\begin{equation}\label{quantity}
\frac{(d-p-1)!}{(d-2n-1)!} \: e_{ca_1...a_p}
\end{equation}
and is calculated in the Appendix.

Let $n^1,..,n^p$ be ortho-normal vectors that span the normal space.
The one free index intersection volume form is defined by
\begin{equation}\label{volume element one index}
\tilde{e}_c= \prod_{i=1}^p(n^i \cdot n^i) \: \:
(n^1)^{a_1}...(n^p)^{a_p} e_{a_1..a_pc}
\end{equation}
Note the difference in the position of the free index $c$ from the
previous formula.

If we define the matrix of components
\begin{equation}\label{1}
u^j_i:=u^a_{i0} n_a^j,
\end{equation}
expanding the vectors $u_{i0}$ in (\ref{lastbefore}) in the
ortho-normal basis we have
\begin{equation}\label{2}
u^{a_1}_{10}..u^{a_p}_{p0} e_{a_1..a_pc}=\det(u^j_i) \tilde{e}_c,
\end{equation}
so finally
\begin{equation} \label{lastbefore2}
({\cal E}^n_{(p)})_c =\frac{n!}{(n-p)!} \frac{(d-p-1)!}{(d-2n-1)!}
\: (-1)^{n-p}(-1)^{p(p-1)/2} \ 2^p  \:
  \det(u_i^j) \:
  \int_{s_{01..p}} d^pt \: \:
(u(t)^2)^{n-p} \: \:  \tilde{e}_c.
\end{equation}
Substituting this in (\ref{Thejunction}) and reinstating $l$, we
get:
\begin{proposition}
The junction condition for the simplicial $p$-intersection is:
\begin{align}\label{pJunction}
(T_{d,01 \dots p})^a_b = & \Lambda_{d,01 \dots p} \,
\delta^a_b,\\\nonumber \Lambda_{d,01 \dots p} = &
\sum_{n=p}^{[(d-1)/2]} \frac{\beta_n}{l^{2n-p}} (-1)^{n-p+1}
\frac{n!}{(n-p)!} \frac{(d-p-1)!}{(d-2n-1)!} \: 2^{p-1} \:
\det(u_i^j) \: \int_{s_{01..p}} d^pt \: \: (u(t)^2)^{n-p}.
\end{align} where $T_{d,01 \dots p}$ is the energy-momentum
tensor on the intersection.
\end{proposition}


Before proceeding let us first look at the Einstein case where only
$\beta_1$ is non zero. AdS and dS spacetimes correspond to the
vector $u$ being space- and time-like respectively so we find that
the bulk cosmological constant is
\begin{equation}\label{}
V_d=-\Lambda_d=\mp \beta_1 \frac{(d-1)(d-2)}{2l^2}
\end{equation}
which is the standard formula with beta related to the Newton's
constant $G$ by $\beta_1=(8\pi G)^{-1}$. The tension of a
hypersurface in Einstein gravity reads
\begin{equation}\label{}
V_{d,10}=-\Lambda_{d,10}=(d-2)\frac{\beta_1}{l}  (u_1-u_0)^a n_a
\end{equation}
where $n^a$ is the normal vector on the hypersurface $\{10\}$.
Applying this to the geometry of the three-way intersection
discussed in section \ref{3way} we find
\begin{equation}\label{einstein hypersurface}
V_{d,10}=(d-2)\frac{\beta_1}{l}
 \big(\sin(\phi_1-\theta_0)-\sin(\phi_0-\theta_0)\big)=(d-2)\frac{\beta_1}{l}\ 2
\sin(\phi_1-\theta_0)
\end{equation}
where $u_i=(\cos \phi_i,\sin\phi_i)$ and $n_i=(-\sin \theta_i,\cos
\theta_i)$, and the positions of the hypersurfaces $\{10\}, \
\{21\},\ \{02\}$ are $\theta_0,\ \theta_1,\ \theta_2$ respectively.
We have labeled the regions by 0,1,2 instead of 1,2,3 as in section
\ref{3way}. We have applied the continuity conditions
(\ref{continuity conditions in angles}) to get the l.h.s. of
(\ref{einstein hypersurface}). Then the tension is positive if
$\phi_1-\theta_0$ is greater than zero and smaller than $\pi$. In a
similar fashion it is possible for all three to have positive
tensions. In particular, the tensions become equal (and positive) in
the symmetric case where the vectors $u$ are symmetrically arranged
and so are the hypersurfaces, with the directions of the $u$'s lying
in between the hypersurfaces at $\pi/3$ angle from them. This setup
on AdS background has been studied in the past, see e.g.
\cite{Csaki:1999mz}. Below we will use the symmetrically arranged
vectors $u$ in the case of general co-dimension to show that
$\det(u^j_i)$ in (\ref{pJunction}) is always positive. For special
Lovelock gravities we will see that the tensions of the
intersections are all positive.
\\


Something interesting about the contributions of the individual
Euler terms, that is, about the value of $\Lambda_{d,01 \dots p}$
when a single such term is considered or contributes, is that it
never vanishes.
%
%
\begin{proposition}\label{genprop}
Each term in (\ref{pJunction}) can not vanish unless $\beta_n$ is
zero. (The terms can possibly cancel among themselves).
\end{proposition}
\emph{Proof:} First recall:
\begin{gather*}
u_i^j = \left(%
\begin{array}{ccc}
  u_{10}^{(1)} & \cdots & u_{10}^{(p)} \\
  \vdots & \ddots & \vdots \\
  u_{p0}^{(p)} & \cdots & u_{p0}^{(p)} \\
\end{array}%
\right)
\end{gather*}
and each vector $u_{i0}$ is proportional to the normal vector of the
hypersurface $\{0i\}$. If the determinant of $u_i^j$ is zero then
the vectors $u_{i0}$ are not linearly independent. That is, they
can't span the $p$-dimensional normal space of the codimension $p$
simplicial intersection so the configuration degenerates to a lower
co-dimension intersection.

Also since
\begin{gather*}
u(t) =\sum_{i=0}^p t^i u_i= u_0+\sum_{i=1}^p t^i u_{i0}
\end{gather*} and $u_i$'s are spacelike vectors then $u(t)^2\geq 0$.
But $u_{i0}$ are linearly independent space-like vectors which span
the normal space and $u_0$ a spacelike vector on it, that is, $u(t)$
cannot be zero everywhere on the $p$-simplex. So the integral in
(\ref{pJunction}) does not vanish.$\Box$

Now define
\begin{equation}\label{poly}
P_{d,p}(x):=2^{p-1}\sum_{n=p}^{[(d-1)/2]} \frac{\beta_n}{l^{2n-p}}
(-1)^{n-p+1} \frac{n!}{(n-p)!} \frac{(d-p-1)!}{(d-2n-1)!} \: x^{n-p}
\end{equation}
where the dependence of $P_{d,p}$ on $\beta$'s and $l$ is
suppressed. ${\Lambda_{d,01 \dots p}}$ then reads
\begin{gather}\label{lambdagenprop2}
\Lambda_{d,01 \dots p}=\det(u^j_i) \int_{s_{01..p}} d^p t \
P_{d,p}\left(u(t)^2\right)
\end{gather}
We have the following
\begin{proposition}\label{genprop2}
A sufficient condition for $\Lambda_{d,01 \dots p} \neq 0$ is
$P_{d,p}(x)
>0$ (or $<0$) for $0 < x < 1$.
\end{proposition}
\emph{Proof:} Since, $|u(t)| \leq \sum_{i=0}^p t^i |u_{i}| =
\sum_{i=0}^p t^i =1$, by $|u_{i}|=1$ with $|u|=\sqrt{u^2}$, the
proposition is clear for $x=u(t)^2$ and $0 \leq x \leq 1$. Also if
$u(t)^2=0$ then $0=u(t)=\sum_{i=0}^p t^i u_i$, so by linear
independence of the $t^i$ we get that all $u_i=0$; this happens only
at one point on the simplex. On a similar basis, if $u(t)^2=1$ then
all vectors $u_i$ must be equal; $u(t)^2=1$ happens only at the
$p+1$ points, the 0-dimensional faces of the simplex. So for the
integral (\ref{lambdagenprop2}) one may take $0 <x<1$ and if
$P_{d,p}>0$ (or $<0$) in this region the integral does not
vanish.$\Box$
\\


An interesing case we can study is Chamseddine's Chern-Simons theory
with AdS gauge group, in $d=$~odd
~\cite{Chamseddine}\cite{dcblackholes}. It is a Chern-Simons theory
from an Euler density in $d+1=$~even dimensions with tangent space
being AdS instead of Minkowski. This Chern-Simons theory is
classically equivalent to a Lovelock gravity\footnote{By considering
intersections in Chamseddine's theory we go through a curious kind
of cycle- Chern-Simons (gauge theory) $\to$ Lovelock Gravity $\to$
Chern-Simons (intersection terms). Whether there is anything deep
behind this or just coincidence, we do not know.}, existing in
$d=$~odd, with coefficients
\begin{equation}
\beta^C_n=\kappa (d-2n)! \frac{(\pm 1)^{n+1}\lambda^{2n-d}}{d-2n}
\begin{pmatrix}
k \\ n
\end{pmatrix}=\kappa (\pm 1)^{n+1}\lambda^{2n-d}(d-2n-1)! \,
\begin{pmatrix}
k \\ n
\end{pmatrix}
\end{equation}
where $n=0,...,k$ with $k=\frac{d-1}{2}=[\frac{d-1}{2}]$ and the
minus (plus) sign corresponds to the dS (AdS) group case and
$\lambda$ is the dS (AdS) gauge group length parameter and $\kappa$
a dimension-less constant. The factor $(d-2n)!$ comes from our
definition of the Euler terms in (\ref{genLove}) compared to the
definition in the references.

It is easy to see that the bulk equations of motion for our AdS
background (\ref{AdScurve}) implies $l^2=\lambda^2$. This is the
vacuum solution of the theory.
As both variables are assumed positive we have $\lambda=l$. Using
the formula for $\beta^C_n$ above in (\ref{poly}) and redefining the
summed index as $n-p=m$ we see that
\begin{eqnarray}\label{cpoly}
&& P_{d,p}(x)=- \kappa 2^{p-1}  l^{p-d} \frac{k!(2k-p)!}{(k-p)!}
\sum_{m=0}^{k-p} \left(\pm  x\right)^m \frac{(k-p)!}{m!(k-p-m)!}=
\\\nonumber && =- \kappa 2^{p-1}
l^{p-d} \frac{k!(2k-p)!}{(k-p)!} \left(1 -x \right)^{k-p} \quad ,
\quad k:=\frac{d-1}{2} \nonumber
\end{eqnarray}

So we obtain the following formula for the co-dimension $p$ membrane
embedded at the intersection of the regions labeled by $0,1,
\dots,p$ in Chamseddine's theory
\begin{equation}\label{chaminterlambda}
\Lambda_{d,01 \dots p}=- \kappa 2^{p-1}  l^{p-d}
\frac{k!(2k-p)!}{(k-p)!} \: \det(u_i^j) \: \int_{s_{01..p}}d^pt
\left(1 - u(t)^2 \right)^{k-p}
\end{equation}

All these $\Lambda$'s are non-zero: as the polynomial does not
change sign for $0 < x < 1$ so we see from Proposition
\ref{genprop2} that (for Chamseddine's theory with AdS group)
$\Lambda_{d,01 \dots p} \neq 0$.

Let the vectors $u_0, \dots, u_p$ be symmetrically arranged in the
normal space forming a symmetric hedgehog. This is discussed in
Appendix \ref{higherthan2} where formulas for $\det(u^j_i)$ and
$u(t)^2$ are obtained. One finds
\begin{equation}\label{symchaminterlambda}
\Lambda_{d,p}=- \kappa l^{p-d} \sqrt2 (2\sqrt{6})^{p-1}
\frac{k!(2k-p)!}{(k-p)!} \left(1+\frac{1}{k}\right)^{k-p/2}
\int_{s_{p}}d^pt \: \{1 - \sum_{i=0}^p t_i^2\}^{k-p}
\end{equation}
where $s_p$ is any $p$-simplex, as by symmetry the tensions of all
$(d-p)$-dimensional membranes in the configuration are the same. One
could say that $ V_{d,p} = -\Lambda_{d,p}$ is the tension of {\it
the} (maximally symmetric) co-dimension $p$ membrane in the vacuum
of Chamseddine's gravity; it has been emphasized at the introduction
that these membranes are embedded in spacetime without causing
singularities or changing its topology, for $p=1, \dots,k$. The
tensions in this formula depend only on the dimensionless $\kappa$,
the length $l$, and the dimensions $d$ and $p$ and they are all
positive.

Let us now turn to $d=$~even. Consider the following Lagrangian in
$d=2k+2$ dimensions defined as
\begin{eqnarray}\label{BI}
&& \kappa f((\Omega \pm \frac{1}{\lambda^2} E \wedge E)^{k+1})=
\\\nonumber && =\kappa \sum_{n=0}^{k} \frac{(k+1)!}{n!(k+1-n)!}(\pm
1)^{k+1-n} \lambda^{2n-2k-2} f(\Omega^n E^{2k+2-2n})+ \kappa
f(\Omega^{k+1}) \nonumber
\end{eqnarray}
For the $+$ ($-$) sign choice the constant curvature vacuum solution
is an AdS (dS) spacetime with curvature proportional to
$\lambda^{-2}$ where $\lambda$ is a length parameter; we will call
(\ref{BI}) as AdS and dS Born-Infeld theories respectively~\cite{BI,
dcblackholes}. $\kappa$ is again a dimension-less parameter.

The last term is topological (exact form locally) and drops out of
the equations of motion. So using (\ref{volform}) and the general
definition of the Lovelock Lagrangian (\ref{genLove}) we find
\begin{equation}
\beta_n^{BI}=\kappa(\pm 1)^{k+1-n} \lambda^{2n-d} d (d-2n-1)!
\begin{pmatrix} k \\ n \end{pmatrix}
\end{equation}
where again $k=d/2-1=[\frac{d-1}{2}]$, $n=0,...,k$. These
coefficients are similar to $\beta_n^C$'s so one may say that
Born-Infeld theory is the analogue to Chamseddine theory in
$d=$~even.

Again the bulk equations of motion for our AdS background
(\ref{AdScurve}) give that $\lambda^2=l^2$ for the AdS Born-Infeld
theory. Putting the AdS $\beta_n^{BI}$'s into (\ref{poly}) we have
\begin{equation}\label{bipoly}
P_{d,p}(x)=-\kappa 2^{p-1}  d\frac{k!(d-p-1)!}{(k-p)!} l^{p-d}
\left(1 - x \right)^{k-p}
\end{equation}
From this we obtain formulas similar to (\ref{chaminterlambda}) and
(\ref{symchaminterlambda}), and from Proposition \ref{genprop2} we
have
that for (AdS Born-Infeld theory) all $\Lambda_{d,01\dots p}$ are
non-zero.


Note that the results for non-vanishing simplicial intersection's
energy tensors are due to the high symmetry of the system: the bulk
regions are portions of the same, highly symmetric spacetime, AdS,
and the gravity theories have an AdS with a given radius as the
single vacuum solution. In general, vanishing (simplicial)
intersection's tensor does {\it not} imply degeneration of the
intersection, that is, the connection can be discontinuous at the
hypersurfaces. On the other hand, it is an interesting fact that the
high symmetry of the background and of the theory makes all these
intersection energy tensors (tensions of the embedded membranes)
strictly non-zero.

\begin{center} *** \end{center}

In $d=$~even it is easy to see why the polynomials get these summed
expressions (\ref{BI}) and in turn by the similarity of the
coefficients, to see why in Chamseddine's Lagrangian expressions get
simplified too. In fact the simplicity has nothing to do with the
AdS background we mainly use in this work: according to our
discussion in ~\cite{Gravanis-04} the simplicial intersection
Lagrangians are generated by expanding the polynomial
\begin{equation}\label{etabi}
\eta_{BI}=\kappa f\left((\Omega_F \pm \lambda^{-2} E(t) \wedge
E(t))^{k+1}\right)=\kappa f\left((d_t \omega +\Omega(t) \pm
\lambda^{-2} E(t) \wedge E(t))^{k+1}\right).
\end{equation}
The intersection Lagrangians read ~\cite{Gravanis-04}
\begin{eqnarray}\label{etaBIp}
&& \int_{s_{01..p}} \eta_{BI} = \kappa (-1)^{p(p-1)/2}
\frac{(k+1)!}{(k+1-p)!} \times \\\nonumber && \times
\int_{s_{01..p}} d^p t
f\left((\omega_1-\omega_0)..(\omega_p-\omega_0) [\Omega(t) \pm
\lambda^{-2} E \wedge E]^{k+1-p}\right)
\end{eqnarray}
from which the equations of motion (junction conditions) are
obtained by merely varying with respect to the frame $E$, as the
variation with respect to the connection vanishes under the zero
torsion condition for the frame on each bulk
region~\cite{Gravanis-04}. For AdS backgrounds and going through the
steps that lead to (\ref{pJunction}) we can show that (\ref{etaBIp})
leads to (\ref{bipoly}). (\ref{etaBIp}) can be applied to more
general backgrounds such as the asymptotically AdS black holes of
these theories, see
e.g.~\cite{BI}\cite{dcblackholes}\cite{Crisostomo}, which will
support less trivial energy tensors and time evolution at the
intersection hypersurfaces.

\section{Non-simplicial intersections and AdS
boundary}\label{non_simp_Section}

We now return to a co-dimension 2 intersection. Let us now see what
happens if there are four or more hypersurfaces intersecting. We
have bulk regions $i =1,\dots,m$ with hypersurfaces given by the
configuration of angles: $\theta_1,\dots,\theta_m$. We label the
intersection as $I$.

The metric continuity condition $(u_i-u_{i+1}) \cdot x_i=0$ for the
$i$-th hypersurface gives
\begin{equation}
\cos(\theta_i-\phi_i)=\cos(\theta_i-\phi_{i+1})
\end{equation}
writing $x_i=\rho_i(\cos\theta_i,\sin\theta_i)$ where $(\rho_i,
\theta_i)$ is the position of the $i$-th hypersurface on plane. One
solution of this equation says that $\phi_i-\phi_{i+1}$ is integer
multiple of $2\pi$ which is rejected as implying that $u_i=u_{i+1}$
which would make the connection continuous there by
(\ref{foundconnection}). The other is
$\theta_i-\phi_{i+1}=-(\theta_i-\phi_i)+2\pi\nu_i$ or
\begin{equation}\label{matchpositions}
\theta_i=\frac{1}{2}(\phi_i+\phi_{i+1})+\nu_i \pi
\end{equation}
$i=1,\dots,m$ with the convention $\phi_{m+1}=\phi_1$. $\nu_i$'s are
integers. there is a discontinuity $u_i \neq u_{i+1}$ which implies
also the discontinuity $\omega_i\neq\omega_{i+1}$ of the connection,
from the formula (\ref{foundconnection}).

Now one finds
\begin{eqnarray}
&& \sum_{j=1}^{i-1} (-1)^j\theta_j +\sum_{j=i}^{m}(-1)^{m-j}\theta_j
=(-1)^i \frac{(-1)^m-1}{2} \phi_i + \bar\nu_i \pi
\end{eqnarray}
where $\bar \nu_i= \sum_{j=1}^{i-1} (-1)^j\nu_j
+\sum_{j=i}^{m}(-1)^{m-j}\nu_j$.

For $m=$ even one finds for all $i$ the single expression
\begin{equation}\label{eventheta}
\theta_1-\theta_2+\theta_3-\theta_4+\dots+\theta_{m-1}-\theta_m=-\bar\nu
\pi
\end{equation}
where all $\bar\nu_i$'s are equal and denoted $\bar\nu$. The angles
$\phi_i$ drop out. We have chosen $0 \leq \theta_1 < \theta_2 <
\dots < \theta_m<2\pi$. It is not hard to see that the above
equation makes sense only for $\bar\nu=1$. So in this case we can't
put the discontinuity hypersurfaces anywhere we like, without making
the metric discontinuous. So $\phi_i$'s can't be expressed in terms
of the positions of the hypersurfaces. The tension in $I$ is only a
function
of the bulk regions data $\phi_i$. 
The intersection behaves rather as part of the background. The same
happens to the analogous situation when we study collisions.

For an $m=$~odd number of hypersurfaces we have from the above
formula
\begin{gather}
\label{uangle} \sum_{j=1}^{i-1} (-1)^j\theta_j
-\sum_{j=i}^{m}(-1)^{j}\theta_j
=(-1)^{i-1} \phi_i+\bar \nu_i \pi.
\end{gather}
We then derive:
\begin{gather}
(-1)^{i}(\phi_{i+1}-\phi_{i}) = 2\sum_{j=1}^{i-1} (-1)^j\theta_j-
2\sum_{j=i+1}^{m}(-1)^{j}\theta_j-(\bar\nu_{i+1}+\bar\nu_i)\pi
\label{thetaphi}.
\end{gather}


We need the junction conditions for a non-simplicial intersection.
It is worthwhile digressing to explain a bit the abstract approach
of ref.~\cite{Gravanis-04} which allows us at once to write down the
answer. Below we give only a sketch of the method. For a full
account the reader should consult~\cite{Gravanis-04}.

The intersection Lagrangians are obtained by expanding a polynomial
\begin{gather}\label{Secondary_Polynomial}
\eta  = (d_t \omega(t)+\Omega(t))^{a_1 \dots a_{2n}} \wedge
e_{a_1\dots a_{2n}}.
\end{gather}
The structure of the Lagrangian at the intersection $\{123\}$ of the
hypersurfaces $\{12\}$, $\{23\}$, $\{31\}$, separating three bulk
regions, given by
\begin{equation}\label{L123}
\mathcal{L}_{123}=\int_{s_{123}} \eta
\end{equation} is a result of
the simplex boundary rule
\begin{equation}\label{sboundary}
\partial s_{123}=s_{23}-s_{13}+s_{12}.
\end{equation}
See Appendix \ref{appsimplex} for the general definition of the
simplex and the associated boundary operator. The form $\eta$ is a
generalized Lagrangian, an example of which we used in
(\ref{etabi}). $\eta$ generates the intersection Lagrangians
according a rule like (\ref{L123}) by integrating over $t$. Here $t$
is the co-ordinate on the simplex. More generally $t$ is the
co-ordinate on a chain which is dual to the intersection in the
following sense.

Consider a non-simplicial intersection, for example of four
hypersurfaces $\{12\}$ ,$\{23\}$, $\{34\}$, $\{41\}$, separating
four bulk regions. We consider a 3-dimensional simplex with vertices
labeled by 1,..,4. The Lagrangian at the intersection is constructed
by finding a chain\footnote{A chain or $p$-chain is, for our
purposes, a linear combination of $p$-dimensional simplices with
coefficients integer or rational numbers.} $c$ on that simplex such
that
\begin{equation}\label{boundaryc}
\partial c=s_{12}+s_{23}+s_{34}+s_{41}
\end{equation}
where the r.h.s. reflects the arrangement of the hypersurfaces on
the normal plane of their intersection.

By (\ref{sboundary}) it is easy to see that such a chain is
\begin{equation}
c=s_{123}+s_{134}
\end{equation}
where the boundary operator acts linearly; $c$ is not unique,
different $c$'s obeying (\ref{boundaryc}) differ by a chain which is
itself a boundary. Then the Lagrangian, given by
\begin{equation}\label{Lc}
\int_c \eta \end{equation} is
\begin{equation}
\mathcal{L}_{123}+\mathcal{L}_{134}.
\end{equation}
As $c$ is not unique, this Lagrangian is not unique either. If $c'$
is another chain satisfying (\ref{boundaryc}) then $c'=c+\partial
\sigma$ for some chain $\sigma$, so by Stokes theorem the
Lagrangians corresponding to them are related by $\int_{c'} \eta=
\int_c \eta+ \int_{\sigma} d_t\eta$. It is a special property of the
Polynomials (\ref{Secondary_Polynomial}) that the pull back of
$d_t\eta + d_x \eta$ onto the $d+1$ dimensional space $\sigma
\times\,$(intersection) vanishes. So Lagrangians constructed by
different chains $c$ differ only by exact forms
\begin{equation*}
\int_c \eta = \int_{c'}\eta + d\int_{\sigma} \eta.
\end{equation*}

It easy to construct now the Lagrangian for the non-simplicial
intersection of $m$ hypersurfaces which reads ${\cal L}_{I} = {\cal
L}_{123} +{\cal L}_{134}+\cdots+{\cal L}_{1,m-1, m}$. This gives
\begin{gather}\label{sumsines}
({T}_{I})^a_b = -2(d-4)(d-3)\beta_2\, \Delta\,
\delta^a_b,\\\nonumber \Delta = \sum_{i=1}^m
\sin(\phi_{i+1}-\phi_{i})
\end{gather}
with $\phi_{m+1} \equiv \phi_1$. Using (\ref{thetaphi}), we can
express $\Delta$ purely in terms of the configuration:
\begin{gather}
\Delta = \sum_i \sin\left(2\sum_{j=1}^{i-1} (-1)^{i-j}\theta_j
-2\sum_{j=i+1}^{m}(-1)^{i-j}\theta_j-(-1)^i
(\bar\nu_{i+1}+\bar\nu_i)\pi\right).
\end{gather}
The solution $\phi_i=\phi_{i+1}$ is trivial so the terms in bracket
can not vanish individually. However, there are more degrees of
freedom than for the three-way intersection. There should be
non-trivial zeroes of $\Delta$. The simplest 3-way planar
intersection (section \ref{intersections}) in AdS background will
have singular matter at the intersection. The intersection of a
higher odd number of branes may or may not, depending on the
geometry.


We now point out an interesting relation between the limit $m \to
\infty$ of the number of intersecting hypersurfaces and the boundary
of AdS. In the example of a non-simplicial co-dimension 2
intersection of $m$ hypersurfaces
let the vectors $u_i$ be arranged symmetrically by
\begin{equation}\label{symphi}
\phi_i=(i-1) \,\frac{2\pi}{m} \: , \: i=1,2, \dots,m
\end{equation}
From (\ref{matchpositions}) and taking $\nu_1= \dots \nu_{m-1}=0$
and $\nu_m=1$, which is also consistent with the constraint
$\bar\nu=1$ in (\ref{eventheta}) for $m$=even, we find
\begin{eqnarray}\label{symtheta}
&& \theta_{i|i\neq m}=\left(i-\frac{1}{2}\right) \frac{2\pi}{m}
\\\nonumber
&& \theta_m=\frac{(m-1)}{m}\pi+\pi=\left(m-\frac{1}{2}\right)
\frac{2\pi}{m} \nonumber
\end{eqnarray}
So the direction of the $u$ vector of every bulk region is in
between of the directions of the hypersurfaces bounding that region.

From the $\phi_i$'s and $\theta_i$'s we find that the metric is
given by $g_{\mu\nu}=\eta_{\mu\nu} (C+\frac{1}{l}\rho
g_m(\theta))^{-2}$ where $\rho$ is the radial variable on the normal
plane, $C>0$ is a constant we usually set to 1, and
\begin{equation}
g_m(\theta)= \left\{ \cos\left(\theta-i\frac{2\pi}{m}\right) \: , \:
-\frac{\pi}{m}+i\frac{2\pi}{m} \leq \theta \leq
\frac{\pi}{m}+i\frac{2\pi}{m} \: , \: i=0,1, \dots, m-1 \right.
\end{equation}
It is continuous and $2\pi$-periodic in the $2\pi$-periodic variable
$\theta$. In fact the function repeats the same values in every
region: at all hypersurfaces has the value $\cos(\frac{\pi}{m})$ and
approaches the value 1 in the middle of the interval; it is a copy
of
\begin{equation}
\cos\theta \: , \: \theta \in [-\frac{\pi}{m},\frac{\pi}{m}]
\end{equation}
for $m$ times. So the interior of the bulk regions is a copy of that
piece of the AdS spacetime with radius $l$, a $1/m$ of the whole. In
particular we have that
\begin{equation}
\cos(\frac{\pi}{m}) \leq g_m(\theta) \leq 1 \quad, \quad \forall  \:
\theta \in [0,2\pi]
\end{equation}
If we take the limit $m \to \infty$ the function $g_m(\theta)$
approaches the constant value 1. In this limit the metric of the
spacetime becomes
\begin{equation}\label{largemmetric}
ds^2_{m\to\infty}=\left(C+\frac{1}{l}\rho\right)^{-2}
(d\rho^2+\rho^2d\theta^2+\eta_{\alpha\beta}dx^\alpha dx^\beta)
\end{equation}
where $x^\alpha$ are the coordinates parallel to the co-dimension 2
intersection.
The curvature 2-form is calculated to be
\begin{equation}\label{largemcurva}
\Omega^{ab}=\delta^a_i\frac{1}{l^2}\left(1+C\frac{l}{\rho} \right)
Q^{i}_{j} E^j \wedge E^b-\delta^b_i
\frac{1}{l^2}\left(1+C\frac{l}{\rho} \right) Q^{i}_{j} E^j \wedge
E^a-\frac{1}{l^2} E^a \wedge E^b
\end{equation}
where $i,j=1,2$ are indices of the Cartesian coordinates on
$(\rho,\theta)$ plane and
$Q^{i}_{j}=\delta^{ij}-\frac{x^ix^j}{\rho^2}$ projection operator on
it, and the Ricci scalar is
\begin{equation}
R=2(d-1)\frac{1}{l^2}\left(1+C\frac{l}{\rho}
\right)-d(d-1)\frac{1}{l^2}
\end{equation}
For $C \neq 0$ the space develops a curvature singularity.

The curvature singularity can actually be removed if the
intersection is located within the boundary of each AdS bulk region.
The constant $C$, taken to be the same for all regions, restricts
the coordinates via $C+u_i \cdot x/l>0$ for the $i$-th region. Call
$A_i$ the space defined by this inequality. If we set ($C=0)$ we
have that the metric in $A_i$ is
\begin{equation}\label{C=0}
ds^2_{i}=\frac{l^2}{(u_i \cdot x)^2} \: \eta_{\mu\nu}
dx^{\mu}dx^{\nu}
\end{equation}
with $u_i \cdot x>0$. We want to include the space $u_i \cdot x=0$
in $A_i$ i.e. to consider the closure $\bar A_i$ of the open $A_i$.
The metric (\ref{C=0}) does not extend over the boundary of this
space and it is given a meaning along the lines of Penrose's
conformal compactification. One may multiply this metric with a
function $f$ with a first order zero at the points $x$ with $u_i
\cdot x=0$, to get a metric $d \tilde s^2_i=f^2 ds^2_i$ which
extends to the boundary $u_i \cdot x=0$ of $A_i$ and defines a
metric $d\tilde s^2_{ib}$ in it; the function $f$ is arbitrarily
chosen in $\bar A_i$, as long as it has a first order zero at the
boundary. As there is no natural choice of $f$, the coefficient of
the zero is arbitrary and the metric $d\tilde s^2_{ib}$ is only
well-defined up to conformal transformations.
$A_i$ is a part of the AdS spacetime, the patch covered in Poincare
coordinates which we have used to write the AdS metric in
(\ref{C=0}) and $u_i \cdot x=0$ is a part of the AdS boundary. The
boundary has the topology of a sphere times the real line: ${\bf
S}^{d-1} \times {\bf R}$ ~\cite{Hawking-73}\cite{Witten:1998qj}.

So let $C=0$ in the bulk regions so that the $i$-th region is a
subspace of $A_i$. The intersection is located at a common
co-dimension 2 subset of the boundary of all $A_i$'s; it is given by
$\rho=0$ in each one of them.
When this is the case the infinite $m$ metric reads
\begin{equation}\label{largemmetric2}
ds^2_{m\to\infty}=\frac{l^2}{\rho^2}
(d\rho^2+\rho^2d\theta^2+\eta_{\alpha\beta}dx^\alpha
dx^\beta)=\frac{l^2}{\rho^2} (d\rho^2+\eta_{\alpha\beta}dx^\alpha
dx^\beta)+l^2 d\theta^2
\end{equation}
which is nothing but a $(d-1)$-dimensional AdS times a circle with
radius $l$: $AdS_{d-1} \times {\bf S}^1$. That is, a dimension gets
compactified and $\rho=0$ becomes the boundary of an AdS (a single
Poincare patch of an $AdS_{d-1}$). The $AdS_{d-1}$ metric is
conformal (with a constant factor) to that of the
$AdS_d|_{\theta={\rm const.}}$ and it is the AdS living at each
hypersurface, ending at $\rho=0$. The boundary of the limiting
spacetime has topology ${\bf S}^{d-2} \times {\bf S}^1 \times {\bf
R}$.

We have mentioned that because the metric at the boundary of the AdS
is defined up to conformal transformations the energy tensor there
has to be traceless. In our case it is diagonal so it should vanish
identically. Now for finite $m$ the tension at the intersection
(\ref{sumsines}) via (\ref{symphi}) reads
\begin{equation}
2(d-3)(d-4) \beta_2 l^{-2} \cdot m \sin \frac{2\pi}{m}
\end{equation}
Absence of a curvature singularity in the limit $m \to \infty$ is
consistent only with $\beta_2=0$ (or $d \leq 4$) i.e. only Einstein
gravity. This is for the symmetric configuration (\ref{symphi}). On
the other hand take $m$=even and consider the configuration
\begin{equation}
\phi_i=(i-1)\frac{2\pi}{m}+(-1)^i \epsilon \quad , \quad i=1, \dots,
m={\rm even}
\end{equation}
where $\epsilon$ is a constant so that by the metric continuity
condition (\ref{matchpositions}) the positions (\ref{symtheta})
remain unchanged, employing the fact that for $m$=even the positions
$\theta_i$ don't fix completely the $\phi_i$'s. The limiting metric
in this case is
\begin{equation}\label{largemmetric3}
ds^2_{m\to\infty}=\frac{l^2}{\cos^2 \epsilon}\frac{1}{\rho^2}
(d\rho^2+\eta_{\alpha\beta}dx^\alpha dx^\beta)+\frac{l^2}{\cos^2
\epsilon} d\theta^2
\end{equation}
namely just the radii $l$ of (\ref{largemmetric2}) are rescaled.
From (\ref{sumsines}) and finite $m$ one finds for the tension on
the intersection
\begin{equation}
2(d-3)(d-4) \beta_2 l^{-2} \cdot m \sin \frac{2\pi}{m} \cdot
\cos(2\epsilon)
\end{equation}
For conformal matter on the intersection this should vanish, which
happens in Einstein-Gauss-Bonnet theory if and only if
\begin{equation}\label{cos}
\cos^2\epsilon=\frac{1}{2}
\end{equation}
determining completely the limiting metric (\ref{largemmetric3}).

Conversely, note that in $d \geq 5$ Einstein gravity alone could not
completely fix the metric (\ref{largemmetric3}). It is the higher
Lovelock term which can reach the co-dimension 2 sub-manifold
$\rho=0$ and fix the metric. In fact one can prove the following
\begin{proposition}\label{mprop} 
Any configuration converges in the limit $m \to \infty$ to the
family of metrics (\ref{largemmetric3}). In Einstein gravity the
whole family is allowed and the limit is ambiguous. When the
Gauss-Bonnet term is included a single element is picked by
(\ref{cos}).
\end{proposition}
\emph{Proof:} From (\ref{matchpositions}) we have
\begin{equation}\label{difftheta}
\theta_{i+1}-\theta_i=\frac{1}{2}(\phi_{i+2}-\phi_i)+(\nu_{i+1}-\nu_i)\pi
\end{equation}
In the $i+1$-th region the argument of the cosine in the
intersections metric (\ref{C=0}) ranges according to
\begin{equation}
\frac{1}{2}(\phi_{i}-\phi_{i+1})+\nu_i \pi \leq \theta-\phi_{i+1}
\leq \frac{1}{2}(\phi_{i+2}-\phi_{i+1})+\nu_{i+1}\pi
\end{equation}
In the limit $m \to \infty$, $(\theta_{i+1}-\theta_i) \to 0$ (or to
$2\pi$ when $i=m$ with $\theta_{m+1}\equiv \theta_1+2\pi$.) From the
first formula we see that the argument of the cosine goes to the
fixed value
\begin{equation}
\frac{1}{2}(\phi_{i}-\phi_{i+1})+\nu_i
\pi=\frac{1}{2}(\phi_{i+2}-\phi_{i+1})+\nu_{i+1}\pi
\end{equation}
up to a possible $2\pi$. From this equality we see that in the limit
$m \to \infty$
\begin{equation}\label{limdiffphi}
\frac{1}{2}(\phi_{i+1}-\phi_{i})=(-1)^i \epsilon+\nu \pi
\end{equation}
for some constant $\epsilon$ for an integer $\nu$. \footnote{
Comparing with (\ref{matchpositions}), this implies $\phi_i \to
\theta_i+(-1)^{i-1}\epsilon$ plus integer multiples of $\pi/2$.}
That is, the limiting metric is given by the one-parameter family of
metrics (\ref{largemmetric3}).

Now the quantity $\Delta$ in (\ref{sumsines}) reads for large $m$
\begin{equation}\label{limdelta}
\Delta= \frac{1}{2} \sum_{i=1}^m \cos(\phi_{i+1}-\phi_i) \sin
2(\theta_i-\theta_{i-1})
\end{equation}
neglecting terms of order $(\theta_i-\theta_{i+1})^2 \sim 1/m^2$ in
the sum. As the sines in (\ref{limdelta}) are of order $1/m$ only
the order 1 part of the cosine's argument matters for large $m$.
This has been identified as $(-1)^i2\epsilon$. $\Delta$ converges to
$2\pi \cos(2\epsilon)$ and the tension on the intersection vanishes
under (\ref{cos}). Put differently, one sets this tension to zero
for any $m$ obtaining relations among $\phi$ whose limit constrained
by (\ref{difftheta}) is given by (\ref{cos}) with
(\ref{limdiffphi}).$\Box$

Summarizing, if the number $m$ of intersecting hypersurfaces
separated by AdS backgrounds goes to infinity, the limiting
spacetime does not have $1/\rho$ curvature singularities if the
intersection is put at the boundary of each AdS region. The
constraint for a traceless energy tensor at the intersection can be
satisfied, as described in Proposition (\ref{mprop}).

\section{Colliding shells and Dominant Energy Condition}\label{DEC_Section}

A collision is described by an intersection with the timelike
coordinate being on the plane of intersection. We take the vectors
$u$ to be timelike, that is, we consider dS spacetime. The three
normal vectors $u_i-u_j$ are spacelike; let
$u_i=(\cosh\zeta_i,\sinh\zeta_i)$ so
$(u_i-u_j)^2=2(\cosh(\zeta_i-\zeta_j)-1) >0$, so the hypersurfaces
are actually timelike. Let the positions of the hypersurfaces be
given by the configuration of rapidities: $\psi_1$, $\psi_2$
$\psi_3$. A general point on a hypersurface is labeled $\tau(
\cosh\psi_i, \sinh \psi_i)$, suppressing the other dimensions.
\footnote{It is clear that the description of collisions in dS is an
analytic continuation of that of intersections in AdS, so some
aspects of intersections can be translated to the collisions.
Consider then a non-simplicial collision, of $m$ hypersurfaces. When
$m=$~even~$\geq 4$ without much thought we get the constraint
\begin{equation}
\psi_1-\psi_2+ \dots +\psi_{m-1}-\psi_m=0
\end{equation}
with the r.h.s. being zero as there is no $2\pi$ periodicity here.
An explicit calculation confirms this. In this case the pressure
$\textsf{p}$ at the intersection is not completely determined by the
rapidities $\psi_i$. One of the $\zeta_i$ must also be specified.}

From the calculation of the previous section we have that the
pressure $\textsf{p}$ in the spacelike collision surface\footnote{
Calculating the energy tensor on a spacelike hypersurface one should
keep in mind that we define the volume element (\ref{volume element
one index}) to be negative for such hypersurfaces so the energy
tensor is minus the value given at (\ref{pJunction}).} is
\begin{equation}\label{dS collisions pressure}
\textsf{p}=2 \sum_{n=2} (-1)^{n}\frac{\beta_n}{l^{2n-2}} n(n-1)
\frac{(d-3)!}{(d-2n-1)!} \det(u^i_j) \int_{s_{012}} d^2t
(u(t)^2)^{n-2}
\end{equation}
One can prove an analogous to Proposition \ref{genprop}. The
reasoning is similar, only now $u(t)^2 \leq 0$, or more specifically
$u(t)^2\le -1$. None of the terms in the sum vanishes alone. So in
general, and in particular for the special Lovelock gravities
described by the Chamseddine and Born-Infeld Lagrangians discussed
above, the pressure $\textsf{p}$ does not vanish. That is in general
intersecting inflationary spacetimes with different timelike
coordinate lead to matter with pressure at their spacelike
intersection in Lovelock gravity.

This explicit example gives us the chance to point out the
following, already clear from the general formulas: in a collision
i.e. intersection of \emph{timelike} hypersurfaces, there is in
general matter appearing at the spacelike collision surface. Viewed
on the normal space, this looks like a collision of particles such
that an instanton may appear at the collision event. Now, the
dominant energy condition~\cite{Hawking-73} is that for all timelike
$\xi^a$, $T_{ab} \xi^a \xi^b \geq 0$ and $T^{ab} \xi_a$ is a
non-spacelike vector, where $T^{ab}$ the energy tensor. This is
clearly violated by the above energy tensor. Thus the dominant
energy condition (DEC) can be violated at collisions in Lovelock
gravity.

In Einstein's theory, the gravitational field equations themselves
can not impose the dominant energy condition. One must also specify
the matter equations of motion or, equivalently, the stress-energy
tensor. For example, the junction conditions allow a space-like
hypersurface with space-like matter.

However as we discussed in~\cite{Gravanis-03}, for colliding shells,
the dominant energy condition at the collision is obeyed if and only
if there is no conical defect. So the dominant energy condition
arises naturally \emph{at the collision} from a condition on the
regularity of the metric. Suppose that we have some matter action to
describe the free shells which respects the dominant energy
condition. When the shells collide, perhaps there could also be some
contact interaction at the collision surface, so in principle, we
could add an interaction term to the matter action. The regularity
of the metric imposes that this interaction term must vanish.

For the higher order Lovelock theories, this condition does not
arise naturally. The junction conditions for the collision surface
are non-trivial. They allow for pressure and momentum localized at
the collision. This pressure and momentum is purely tangential to
the surface. So the collision process will involve something flowing
along the space-like collision surface in violation of the dominant
energy condition. So the higher order Lovelock theories impose no
energy condition on the type of interaction allowed. In general, we
could have a collision where all of the shells are ingoing and
annihilate each other, with the energy flowing away to spatial
infinity along the collision surface. These novelties arise from the
peculiar fact that in the energy exchange
relations~\cite{Gravanis-03} for the collision of shells the purely
stress tensor at the spacelike collision surface contributes with
components normal on the surface.

On the other hand, when we consider the matter component of the
theory it is very natural to impose the DEC, which is interpreted as
that the energy can not flow faster than the speed of light. If the
matter part of the theory is such that the DEC is respected, then
this places a strong restriction on the kinds of geometry which are
allowed. For example, if two maximally symmetric shells collide in
dS, it is impossible to have a single outgoing maximally symmetric
brane in dS bulk. There must be more than one outgoing brane and/or
some disturbance of the bulk. So we have a constraint which is a
kind of selection rule for the allowed collisions due to the higher
order Lovelock terms.

As a last comment, we should note that the other energy conditions
are also violated in general. The dominant condition for a perfect
fluid with energy density $\varrho$ and pressure $\textsf{p}$
reads $\varrho \ge |\textsf{p}|$. As discussed this is not
satisfied in an arbitrary collision of shells in Lovelock gravity
because $\varrho=0$ and $\textsf{p}$ is in general non-zero. The
weak energy condition reads $\varrho \ge 0$ and
$\varrho+\textsf{p} \ge 0$. This is satisfied in the examples we
discussed above if $\textsf{p} \ge 0$.\footnote{ Note that
(regarding the bulk cosmological constant as matter) the dS
background itself satisfies the dominant and weak energy
conditions, and also the null energy condition which simply reads
$\varrho+\textsf{p} \ge 0$. It does not satisfy the strong energy
condition: $\varrho+\textsf{p} \ge 0$ and $\varrho+3\textsf{p} \ge
0$ as $\textsf{p}=-\varrho \le 0$. The AdS background satisfies
the null and strong energy condition but it does not agree with
the weak and the dominant energy conditions.} This is certainly
not the case in general: for example if we calculate (\ref{dS
collisions pressure}) for the case of Chamseddine gravity we get
\begin{equation}\label{dS collisions pressure Chamseddine}
\textsf{p}=(-1)^{k+1} \cdot \frac{1}{2} \kappa l^{2-d}
(d-1)(d-3)(d-3)! \ \det(u^i_j) \int_{s_{012}} d^2t \
(-1-u(t)^2)^{k-2}
\end{equation}
where $k=\frac{d-1}{2}$. This is positive if $d=4m-1$, for some
integer $m$, and violates the weak energy condition by being
negative in $d=4m+1$ dimesions.

\section{Dimensionalities of intersections and 4-dimensional
brane universe}\label{dimension_Section}

In $d$ bulk dimensions Lovelock Lagrangian contains terms of $n$-th
power of the curvature, with $d > 2n$, or $n_{\rm max}=[(d-1)/2]$, [
] the integer part. The lowest dimensional intersection is $d-n_{\rm
max}$ or
\begin{equation}
d-\left[\frac{d-1}{2}\right]
\end{equation}
That is, one can't have an intersection of dimension lower that
roughly half the bulk dimensionality, or, for a given intersection
dimensionality the maximum possible bulk dimensions are roughly
twice that. In particular if we are interested in 4-dimensional
sub-manifolds it is easy to see that the available bulk
dimensionalities are $d=5,6,7$.

Consider a spacetime without boundary or with boundary that is
smooth, i.e. the normal direction changes continuously along it. Let
us also insist that spacetime is a (differentiable) manifold. The
metric is assumed non-singular in the sense of being
$C^{1-}$~\cite{Hawking-73}, which in particular means that the first
derivative of the metric may have only finite discontinuities and
remains bounded in general. This excludes conical singularities.
Also, it excludes the general case of orbifolds.

We want to restrict matter in sub-manifolds under these conditions.
Consider a manifold and let finite discontinuities of the first
derivative of the metric occur at hypersurfaces, which in general
intersect. This respects the above conditions. Also, in Lovelock
gravity matter does get localized at (restricted on) the
discontinuities and their intersections. One may say that this is
the only way to get matter restricted in sub-manifolds under the
conditions set in the previous paragraph as an alternative is not
known.

So intersections provide the means to restrict matter in
sub-manifolds of co-dimension 2 or higher in a non-singular
spacetime. Lovelock gravity is in a certain sense a natural
generalization of Einstein theory in $d \geq 5$ dimensions:
(\ref{Lovelockscalar}) is the most general action for the metric
field which produces at most second order field equations under the
condition of zero torsion~\cite{Lovelock-71,Zumino-86}. Under these
conditions and the conditions discussed above, one may say that the
4-dimensional intersections in $d \leq 7$ dimensions exhaust the
list of possibilities in the spirit of the idea to think of our
universe as a subspace of higher dimensional spacetime; one of
course may consider theories of fourth or higher order field
equations, hypersurfaces of arbitrary thickness, conical or other
singularities in the bulk geometry etc, but all these add a very
large number of model depended possibilities in the already not
entirely economical RS-scenario. Another possibility is to assume
that subspaces have their own intrinsic gravity terms, apart from
the ones induced by the bulk. Then one has to invent mechanisms of
how they arise.

Let now the boundary of the spacetime be not smooth, the direction
of its normal vector changes discontinuously crossing hypersurfaces
embedded in the boundary. Then new surface terms should be added in
action involving various angles. In Einstein gravity this has been
analyzed in the past ~\cite{Hayward:1993my}, and has also been used
in the intersecting brane world literature ~\cite{Kaloper:2004cy}.
This kind of action could also be constructed for the general
Lovelock gravity.


\section{Vacuum solutions and higher co-dimension membranes}
\label{high}

The higher order derivative structure of Lovelock gravity allows for
vacuum solutions when the connection is discontinuous at non-null as
well as null hypersurfaces. As in general there are non-trivial
junction conditions at the intersections, higher co-dimension
membranes are allowed to exist in vacuum without deficit angle or
more pathological curvature singularities.

\subsection{Solitonic configurations}\label{solitonsection}

Consider a single hypersurface between the regions labeled by 0 and
1 and a case where
\begin{equation}\label{soliton1}
\Lambda_{d,01}=0
\end{equation}
This is to be satisfied for $\beta_n$'s , $l$ and the $u$'s, along
with the bulk equation of motion (\ref{bulkeq}). Solving
(\ref{soliton1}) for $l$, when it is possible, we obtain an
$l=l(\beta_{n \geq 1},u_i)$. This put in (\ref{bulkeq}) gives the
bulk cosmological constant $\Lambda_{\rm
bulk}=-\frac{1}{2}\beta_0=-\frac{1}{2} \beta_0(\beta_{n \geq
1},u_i)$.

This is a solitonic configuration in the sense that via a
discontinuity in the connection at a hypersurface there exists in
spacetime a self-supported vacuum gravitational field (modulo the
bulk cosmological constant). For Einstein-Gauss-Bonnet theory such a
spacetime was considered, in relation to brane-world problems, in
~\cite{Iglesias:2000iz}.

This kind of solutions (stable or not) existing for some Lovelock
gravities are not related to topological numbers and one obvious
statement is that they are due to the appearance of more than one
delta function (or, to delta function $\times$ zero) in the field
equations in Lovelock gravity in the presence of a hypersurface. A
different way to put it is that they are possible because Lovelock
gravity is on the verge of not having a well defined initial value
problem~\cite{Choquet-Bruhat-88}: as a space-like hypersurface
evolves it is possible to pass through a stage where its extrinsic
curvature jumps without matter being responsible for that. This is
of course not a problem in the solitonic configurations, as they
{\it are} solutions over the whole of time.

Now, when hypersurfaces intersect, in general matter will be
localized at the intersection and so the same can happen when
solitons intersect. Consider first a co-dimension 2 intersection.
When the energy tensor at the co-dimension 2 hypersurface does not
vanish we have a case where a co-dimension 2 matter is standing
alone in spacetime without the appearance of a conical singularity.
By (\ref{pJunction}), define coefficients $c^n_{01 \dots p}$ via
\begin{equation}\label{coeff}
\Lambda_{d,01..p}=\sum_{n=p}^{k}c^n_{01..p} \: \beta_n
\end{equation}
with $k=[\frac{d-1}{2}]$. The dependence of the $c$'s on $d$ is
understood. Let $d \geq 5$ and consider Einstein-Gauss-Bonnet
theory. Three intersecting solitonic configurations of the kind
discussed above means that
$\Lambda_{d,01}=\Lambda_{d,12}=\Lambda_{d,20}=0$ so
\begin{eqnarray}\label{system}
&& c^1_{01} \beta_1+c^2_{01} \beta_2=0
\\\nonumber && c^1_{12} \beta_1+c^2_{12} \beta_2=0
\\\nonumber && c^1_{20} \beta_1+c^2_{20} \beta_2=0
\end{eqnarray}
For $\beta$'s not to be zero the relations have to be linearly
dependent. This is possible: it is adequate to take all three angles
between $u_0,u_1,u_2$ equal, as then $u(t)^2$ will give the same
integral over all 1-simplices. One the obtains a relation for the
couplings constants, that is, is specified a class of the Lovelock
gravities that accommodates such a configuration. Let $u_0=(1,0)$,
$u_1=(-1/2,\sqrt{3}/2)$, $u_2=(-1/2,-\sqrt{3}/2)$. We find
\begin{equation}\label{beta12eq}
-l^{-1} (d-2) \det(u) \, \beta_1+l^{-3} (d-2)(d-3)(d-4) \det(u)
\,\beta_2=0
\end{equation} So for $d \geq 5$ solving for $\beta_2$ we have
by (\ref{coeff})
\begin{equation}
\Lambda_{d,012}=c^2_{012} \beta_2=-3\sqrt{3} \ \beta_1
\end{equation}
where we used also the volume of the 2-simplex
$\int_{s_{012}} d^2 t =\frac{1}{2}.$
As the geometry does not contain a deficit angle (which is not hard
to see employing the metric continuity conditions) we have a
co-dimension 2 surface filled with matter in spacetime without
conical singularities. Moreover the energy density is positive, the
tension on the co-dimension 2 intersection is
$V_{d,012}=-\Lambda_{d,012}=3\sqrt{3} \ \beta_1>0$. We discuss
higher than 2 co-dimension membranes in AdS and the associated
Lovelock gravities in the Appendix \ref{higherthan2}. In general
backgrounds, solitonic solutions are possible with no relations
among the beta couplings. This interesting implication of Lovelock
gravity is discussed in future work.

\subsection{Shock waves}\label{shocksection}

We turn now to the case of shock waves ~\cite{Penrose}. Let a
hypersurface separating a dS region with vector $u_0,\, u_0^2=-1$,
and an AdS region with vector $u_1,\, u_1^2=1$. The hypersurface is
given by the continuity condition $(u_0-u_1) \cdot x=0$. Let the
vector $u_0-u_1$ be null. Then also $u_0 \cdot u_1=0$.
\begin{proposition}
Consider pure Gauss-Bonnet gravity with cosmological constant. Then
a null hypersurface separating a dS and an AdS with the same length
scale $l$ is a shock wave.
\end{proposition}

\emph{Proof} : First, by the bulk equations (\ref{bulkeq}) we see
that if $\beta_n$ is not zero for $n=0,2=$even, the same $l$ can be
a solution for both dS and AdS. Now, let us repeat the equation
(\ref{lastbefore})
\begin{equation} \label{nulllastbefore}
{\cal E}^n_{(p)} =
  (-1)^{n-p}(-1)^{p(p-1)/2} \ 2^p  \
  \int_{s_{01..p}} d^pt \;
u^{a_1}_{10}..u^{a_p}_{p0} \: \: (u(t)^2)^{n-p} \:
\frac{(d-p-1)!}{(d-2n-1)!} \: e_{ca_1...a_p}
\end{equation}
This is still valid even if one (or more) of the $u_{i0}$'s is null.
It is important that we nowhere refer to the intrinsic geometry of
the null hypersurface. If this quantity vanishes there cannot be a
non-zero energy tensor in the null hypersurface.

For Gauss-Bonnet gravity and a null hypersurface discontinuity the
single contribution is $\beta_2$ times (putting the common length
$l$ back)
\begin{equation}
-2 l^{-3}(d-3)(d-4) \int_0^1 dt \ u(t)^2 \ u^a_{10} e_{ca}
\end{equation}
But
\begin{equation}
u(t)^2=-t_0^2+t_1^2=-(1-t)^2+t^2
\end{equation}
taking $t_1=t$, that is
\begin{equation}
\int_0^1 dt \ u(t)^2 =0
\end{equation}
so the energy tensor at the null discontinuity vanishes
identically.$\Box$

Also we have
\begin{proposition}
Let Lovelock gravity be given by a sum of even order Euler terms.
Then a null hypersurface separating a dS and an AdS with the same
length scale $l$ is a shock wave.
\end{proposition}
\emph{Proof} : By the bulk equations (\ref{bulkeq}) we see that if
$\beta_n$ is not zero for $n=$even, the same $l$ can be a solution
for both dS and AdS. For the null hypersurface between dS and AdS
with the same scale we have from before that
$u(t)^2=-(1-t)^2+t^2=2t-1$ so
\begin{equation}\label{integral}
\int_0^1 dt \ (u(t)^2)^{n-1}=\int_0^1 dt \
(2t-1)^{n-1}=\frac{(-1)^{n-1}+1}{2} \, \frac{1}{n}
\end{equation}
so from (\ref{nulllastbefore}) for $p=1$ we have that the energy
tensor at the null hypersurface vanishes for all $n$=even.$\Box$

Consider now the non-simplicial intersection such that four dS and
AdS regions are put alternatively: $u_0=(1,0)$, $u_1=(0,1)$,
$u_2=(-1,0)$, $u_3=(0,-1)$. Let the gravity be pure Gauss-Bonnet.
Then all four hypersurfaces are null and shocks.

The co-dimension 2 hypersurface is spacelike. Its Lagrangian is
\begin{equation}\label{L2}
\mathcal{L}_{012}+\mathcal{L}_{023}
\end{equation}
The intersection is non-null so by (\ref{pJunction}) we get the
energy tensor on it is pure pressure equal to
\begin{equation}
8 \beta_2 \ l^{-2} (d-3)(d-4)
\end{equation}
This is calculated via the determinants of
$\left(\begin{smallmatrix} u_{10} \\u_{20} \end{smallmatrix}\right)$
and $\left(\begin{smallmatrix} u_{20} \\u_{30}
\end{smallmatrix}\right)$ which both equal to 2.

In general, consider the same configuration for a Lovelock gravity
involving all possible even order Euler terms. The Lagrangian is
still given by (\ref{L2}). Over the simplex $s_{012}$ we have
\begin{equation}
u(t)^2=-t_0^2+t_1^2-t_2^2+2 t_0
t_2=t_1^2-(t_0-t_2)^2=(1-2t_0)(1-2t_2)
\end{equation}
using the $u$'s above and that $t_0+t_1+t_2=1$ over this simplex.
There is a similar expression over $s_{023}$. So we have
\begin{equation}
\int_{s_{012}}d^2t \ (u(t)^2)^{n-2}=\int_0^1 dt_0 \int_0^{1-t_0}
dt_2 \left((1-2t_0)(1-2t_2)\right)^{n-2} =\frac{1}{2(n-1)^2}
\end{equation}
where we used formula (\ref{integral}) and that $n$=even. The same
quantity is obtained from the simplex $s_{023}$. Having calculated
the determinants above (equal to 2) we use these to the formula
(\ref{pJunction}) to get for the pressure at the intersection
\begin{equation}
\sum_{n={\rm even} \geq 2} \frac{4n}{n-1}\frac{\beta_n}{l^{2n-2}}
\frac{(d-3)!}{(d-2n-1)!}
\end{equation}
We see then that in a collision of shocks a co-dimension 2 matter is
required to exist at the collision event surface. As noted above
this spacelike matter violates the dominant energy condition, as a
general feature of collisions in Lovelock gravity, here seen in the
case of shock waves.

\paragraph{Acknowledgements}

The main part of this work was done at Kings College London. S.W.
would also like to thank the staff at CECS for discussions during
the revision of this manuscript. S.W. was partially funded by
FONDECYT grant 3060016. The generous support to CECS by Empresas
CMPC is also acknowledged. CECS is a Millennium Science Institute
and is funded in part by grants from Fundaci\'{o}n Andes and the
Tinker Foundation.

\appendix
\numberwithin{equation}{section}

\section{Some manipulations with Kronecker delta}

In this appendix, we derive the quantity given in equation
(\ref{quantity}). We would like to determine the constants
$A(d)_{mn}$ in
\begin{equation}
\frac{(m+n+1)!}{(n+1)!} \delta^{a_1}_{[a_1}..\delta^{a_m}_{a_m}
e_{cb_1...b_m]}= A(d)_{mn} e_{cb_1..b_m}.
\end{equation}
In components this means
\begin{equation}
\frac{(m+n+1)!}{(n+1)!} \delta^{a_1}_{[a_1}..\delta^{a_m}_{a_m}
\epsilon_{cb_1...b_m]c_{n+2}..c_d}= A(d)_{mn}
\epsilon_{cb_1..b_mc_{n+2}...c_d}.
\end{equation}
Contracting with the same epsilon symbol with indices upstairs we
have, using standard formulae (see e.g. the Appendix of
~\cite{Wald-84}),
\begin{equation}
-d! \ A(d)_{mn}=-(d-n-1)!(m+n+1)!
\delta^{a_1}_{[a_1}..\delta^{a_m}_{a_m}\delta^c_c\delta^{b_1}_{b_1}..\delta^{b_n}_{b_n]}.
\end{equation}
It is easy to show that the contracted delta's times $(m+n+1)!$ give
\begin{equation}
\frac{d!}{(d-m-n-1)!}
\end{equation}
so
\begin{equation} \label{A(d)}
A(d)_{mn}=\frac{(d-n-1)!}{(d-m-n-1)!}.
\end{equation}

\section{Variational principle for metric and vielbein}
The action for Lovelock theory with matter is:
\begin{gather*}
{\cal S} = \int_M {\cal L}_{Lovelock}  + \int_M {\cal L}_{mat}.
\end{gather*}
The Euler variation w.r.t. $g^{\mu\nu}$ (neglecting boundary terms)
leads to:
\begin{gather*}
\delta S =\int_M (H_{\mu\nu} - T_{\mu\nu}) \delta g^{\mu\nu}{e},
\end{gather*}
where $H_{\mu\nu}$ is the Lovelock tensor, $T^{\mu\nu}$ the
stress-energy tensor. The volume element ${e}$ is:
\begin{gather*}
{e} = \sqrt{-g}\ dx^1\wedge \cdots\wedge dx^d.
\end{gather*}

These more familiar expressions for the gravitational action
principle are in terms of variation w.r.t. the metric. Since we have
used the vielbein language, it is useful to be able to translate
between the two. The volume element is, in terms of vielbeins:
\begin{gather}
{e} = \frac{1}{d!}\epsilon_{a_1 \dots a_d} E^{a_1} \wedge \cdots
\wedge E^{a_d} = E^{(1)} \wedge \cdots\wedge E^{(d)}.
\end{gather}
We also define
\begin{gather}
{e}_{a_1 \dots a_p} : = \frac{1}{(d-p)!}\epsilon_{a_1 \dots a_d}
E^{a_{p+1}} \wedge \cdots \wedge E^{a_d}
\end{gather}

We shall need these identities:
\begin{align}
E^{c_1\dots c_n} \wedge e_{d_1\dots d_m} & = \frac{m!}{(m-n)!}
\delta^{c_1}_{[d_{m-n+1}}\dots \delta^{c_n}_{d_m} \: e_{d_1 \dots
d_{m-n}]},\label{Ewedgee}\\
\nonumber\\
\delta E^b & =
\delta E^b_\mu E^\mu_c E^c, \label{varyEtog}
\\\nonumber\\
\delta E^b_\mu E^\mu_a & = - E^b_\mu \delta
E^\mu_a,\label{-relation}
\\\nonumber\\
\delta g^{\mu\nu} & =
2\eta^{ab} \delta E_a^\mu E_b^\nu.\label{varygtoE}
\end{align}
The point is that because the $\omega$ equation of motion vanishes
identically, we can use (\ref{varygtoE}) replace metric variations
directly for vielbein variations, $\delta_{g}{\cal L} =
\delta_{g(E)} {\cal L}$. First, we define
\begin{gather}
T_{ab} : = E^\mu_a T_{\mu\nu} E^\nu_b,
\end{gather}

Using (\ref{Ewedgee} - \ref{varygtoE}) and noting that
\begin{gather*}
\delta E^b \wedge E_a = - E^b_\mu \delta E^\mu_a {e},
\end{gather*}
we find that
\begin{gather*}
T_{\mu\nu}\delta g^{\mu\nu}{e} = -2T_b^{\ c} \delta E^b \wedge
{e}_c.
\end{gather*}
The field equations in terms of the vielbeins are:
\begin{gather}
\delta_{E^c} {\cal L}_{\text{Lovelock}} = -2 T^b_c {e}_b.
\end{gather}
\\

If there is singular matter with support on some intersection $I$,
we have a term in the action:
\begin{gather}
\int_I \tilde{\cal L}_{\text{mat}},
\end{gather}
The variation gives the stress-energy tensor on $I$:
\begin{gather}
\delta \tilde{{\cal L}}_{{\rm mat}} \equiv-\tilde{T}_{\mu\nu}\delta
h^{\mu\nu} \tilde{{e}}
\end{gather}
is the energy-momentum tensor on $I$. On the intersection we have an
induced metric $h$ and the corresponding volume element
\begin{gather}
\tilde{e} = \sqrt{|h|}d^{d-p}x.
\end{gather}

The stress-energy tensor will be related to the variation of the
appropriate boundary term in the Lovelock action:
\begin{gather*}
\sum_n \beta_n \int_{I} {\cal L}^n_{(p)}.
\end{gather*}

Let $n^1, \dots , n^p$ be an ordered set of ortho-normal vectors
which spans the space of vectors normal to $I$. In terms of the
vielbeins, the volume element is:
\begin{gather}
{\tilde e} = \prod_{i=1}^p(n^i \cdot n^i) \: \: (n^1)^{a_1}\cdots
(n^p)^{a_p} e_{a_1 \dots a_p}.
\end{gather}
The order of the normal vectors gives the orientation on $I$. The
factor $\prod (n^i \cdot n^i)$ is $\pm 1$ depending on whether $I$
is time-like or space-like.

If we vary the frames tangential to $I$ such that they remain
tangent\footnote{This is sufficient if we vary $\tilde{\cal
L}_{\text{mat}}$ only w.r.t. the induced metric $h$ and not the
position of the intersection. In this paper, we consider only dS/AdS
bulk solutions, where the terms involving $\delta E^a$ not
tangential always vanish anyway.} to $I$, there is a simple
relation:
\begin{gather}
\delta E^a \wedge \tilde{e}_b = E_{b}^\mu \delta E_\mu^a\, \tilde{e}
  \qquad (\delta E^a \text{tangential}).
\end{gather}
Following the same procedure as above, we then derive:
\begin{gather}
\sum_n \beta_n\  \delta_{E^c}\! {\cal L}^n_{(p)} = -2 \tilde{T}^b_c
{e}_b
\end{gather}
It is important to remember this factor of $-1/2$ when relating the
stress-energy tensor to the Euler variation w.r.t. the vielbein.
This has been used in equation (\ref{Thejunction}).

\section{A word on the simplex}\label{appsimplex}

A (Euclidean) $p$-dimensional simplex or $p$-simplex $s_p$ is
defined as $\{t \in \mathbf{R}^{p+1} | \sum_{i=0}^p t^i=1, \, {\rm
all} \, t^i \geq 0\}$. A bit more generally is defined as the set of
points $\sum_{i=0}^p t^i a_i$ with the same conditions for the
$t_i$'s as above, for $a_0,..,a_p$ points in the Euclidean space
$\mathbf{R}^{p+1}$ such that $a_1-a_0,...,a_p-a_0$ are linearly
independent ~\cite{simplexbook}. This reflects nicely the properties
of the vector $u(t)$ encountered in this paper.

A 0-simplex is a point, a 1-simplex is an interval, a 2-simplex is a
triangle, a 3-simplex is a tetrahedron etc. A $k$-dimensional face
of the simplex, designated $s_{i_0..i_k}$, is the subset of $s_p$
such that
\begin{equation}
s_{i_0..i_k}=\{ t \in s_p | t^j = 0, \forall j \neq i_0,\dots i_k \}
\end{equation}
Of course by definition a $k$-dimensional face is itself a
$k$-simplex. It is easy to see that there are
$\left(\begin{smallmatrix} p+1 \\ k+1
\end{smallmatrix}\right)$ $k$-dimensional faces on the $p$-simplex.

Clearly $(k-1)$-simplices are parts of the boundary of the
$k$-simplices. The rule which takes into account orientations is
\begin{equation}
\partial s_{i_0...i_k}=\sum_{r=0}^k (-1)^r s_{i_0...\widehat{i_r}...i_k}
\end{equation}
where hat means that this is index is absent. The symbols
$s_{i_0..i_k}$ are completely anti-symmetric.

\section{Membranes of co-dimension higher than 2 in
AdS bulk}\label{higherthan2}

\subsection{The symmetric hedgehog}

Let the vectors $u$ be symmetrically arranged in the $k$-dimensional
normal space, forming a symmetric hedgehog. In particular the
average position defined by their ends (the barycenter) coincides
with the origin
\begin{equation}
u_0+\cdots +u_k=0
\end{equation}
Also by symmetry all inner products are equal. Call this cosine
$\cos \phi_k$. Taking the square of the above we have
\begin{equation}
(1+k \cos \phi_k )(k+1)=0
\end{equation}
So
\begin{equation}\label{cosine}
\cos \phi_k=-\frac{1}{k}
\end{equation}
Then
\begin{equation}
u(t)^2=\left(\sum_{i=0}^p u_i t_i\right)^2= \sum_{i=0}^p t_i^2+ \cos
\phi_k \sum_{i \neq j} t_i t_j=
\left(1+\frac{1}{k}\right)\sum_{i=0}^p t_i^2-\frac{1}{k}
\end{equation}
where in the last equality we used (\ref{cosine}) and that $( \sum_i
t_i )^2=1$.

The other bit we need is the determinants made out of the vectors
$u_{i0}$. It equals to the volume of an $p$ dimensional
parallelepiped made out of vectors with length
\begin{equation}\label{i0length}
|u_i-u_0|=\sqrt{2-2\cos \phi_k}= \sqrt{2+\frac{2}{k}}
\end{equation}
and angle between any two vectors $u_{i0}$ and $u_{j0}$ given by the
cosine
\begin{equation}
\frac{(u_i-u_0) \cdot (u_j-u_0)}{|u_i-u_0||u_j-u_0|}=\frac{1-\cos
\phi_k}{2(1-\cos \phi_k)}=\frac{1}{2}=\cos60^{o}
\end{equation}
So all ``heights" of the parallelepiped are given by the length in
(\ref{i0length}) times $\sin 60^{o}=\sqrt{3}/2$. The determinant
related to a $p$-simplex face of the $k$-simplex is
\begin{equation}
\det(u_i^j)=\left( \frac{\sqrt{3}}{2} \right)^{p-1}
\left(2+\frac{2}{k} \right)^{p/2}
\end{equation}

\subsection{Gravity}



In the case of Chamseddine and Born-Infeld type of Lovelock gravity
we considered (AdS) vacuum and intersections such that the $u$
vectors are symmetrically arranged. Here we find the Lovelock
gravity such that the tension of all membranes with co-dimension $p
\neq k=[\frac{d-1}{2}]$ is zero.

So $\Lambda_{d,01 \dots p}=\Lambda_{d,p}$ here and for each $p$ we
consider one expression and call the coefficients in (\ref{coeff})
simply $c^n_p$. If all except the co-dimension $k$ intersection
$\Lambda$'s vanish, we have
\begin{eqnarray}
c^1_1 \beta_1+ c^2_1 \beta_2 +  \cdots +c^{k-1}_1 \beta_{k-1}=-c^k_1
\beta_k && \\\nonumber c^2_2 \beta_2+  \cdots +c^{k-1}_2
\beta_{k-1}=-c^k_2 \beta_k &&
\\\nonumber
\vdots \\\nonumber c^{k-1}_{k-1} \beta_{k-1}=-c^k_{k-1} \beta_k
\nonumber
\end{eqnarray}
were we put the last term in the sums in r.h.s. to give it the form
of an upper triangular $(k-1) \times (k-1)$ linear system.

The inverse of the matrix $\mu=(c^n_p)$ of the coefficients is
rather easily calculated by the method of forming a $(k-1) \times
2(k-1)$ matrix by putting a unit matrix on the side of $\mu$ and
adding appropriate multiples of lines to other lines of this big
matrix until in the place of $\mu$ the unit matrix appears; then in
place the unit matrix $\mu^{-1}$ appears. Then we find that the
inverse $\mu^{-1}$ is an upper triangular matrix. The diagonal terms
are $1/c^p_p$ and the upper triangular part is
\begin{equation}
(\mu^{-1})^n_p=-\frac{c^n_p}{c^n_n c^p_p}
\end{equation}
$n=p+1,..,k-1$. So
\begin{equation}\label{newbetas}
\beta_p=-\frac{c^k_p}{c^p_p}\beta_k+\sum_{n=p+1}^{k-1} \frac{c^n_p
}{c^n_n c^p_p}c^k_n \beta_k =
\left(-\tilde{c}^k_p+\sum_{n=p+1}^{k-1} \tilde{c}^n_p \tilde{c}^k_n
\right) \beta_k
\end{equation}
$p=1,..,k-1$, where
\begin{equation}\label{ratios}
\tilde{c}^n_p:=\frac{c^n_p}{c^p_p}=l^{2p-2n} (-1)^{n-p}
\frac{n!}{(n-p)!}  \frac{(d-2p-1)!}{(d-2n-1)!} \int_{s_p} d^p t \,
(u(t)^2)^{n-p}
\end{equation}
These coefficients depend on $d$ and the AdS radius $l$ but also on
a coupling beta which is left arbitrary. In detail, $\beta_0$ is
fixed in terms of a given bulk cosmological constant $\Lambda_{\rm{
bulk}}=-\frac{1}{2}\beta_0$, and by the bulk equations of motion
(\ref{bulkeq}) the coupling $\beta_k$ is fixed in terms of $l$ and
$\Lambda_{\rm{ bulk}}$. Einstein gravity with arbitrary cosmological
constant is the trivial case of these.

\small{


\end{document}